\documentclass[12pt,fleqn]{article}
\usepackage{graphicx,cite,color}

\def\theequation{\thesection.\arabic{equation}}
\newcommand{\newsection}[1]{\section{#1}\setcounter{equation}{0}}
\newcommand{\newappendix}[1]{\section*{#1}\setcounter{equation}{0}}

\def\be{\begin{equation}}
\def\ee{\end{equation}}
\def\bea{\begin{eqnarray}}
\def\eea{\end{eqnarray}}
\def\nnb{\nonumber}
\def\bbuildrel#1_#2^#3{\mathrel{\mathop{\kern 0pt#1}\limits_{#2}^{#3}}}
\def\slash#1{\setbox0=\hbox{$#1$}#1\hskip-\wd0\dimen0=5pt\advance
       \dimen0 by-\ht0\advance\dimen0 by\dp0\lower0.5\dimen0\hbox
         to\wd0{\hss\sl/\/\hss}}

\newcommand{\scs}{\scriptscriptstyle}
\newcommand{\f}{\frac}

\newcommand{\al}{\widetilde{\alpha}_{\mathrm s}}

\begin{document}

\begin{titlepage}

\begin{flushright}
TTP06-24\\
SFB/CCP-06-37\\
IFT-16/2006\\
hep-ph/0609241\\[2cm]
\end{flushright}

\begin{center}
\setlength {\baselineskip}{0.3in} 
{\bf\Large 
  NNLO QCD Corrections to the {\boldmath ${\bar B}\to X_s\gamma$}\\ 
  Matrix Elements Using Interpolation in {\boldmath $m_c$} }\\[2cm]
\setlength {\baselineskip}{0.2in}
{\large  Miko{\l}aj Misiak$^{1,2}$ and Matthias Steinhauser$^3$}\\[5mm]
$^1$~{\it Institute of Theoretical Physics, Warsaw University,\\
         Ho\.za 69, PL-00-681 Warsaw, Poland.}\\[5mm] 
$^2$~{\it Theoretical Physics Division, CERN, CH-1211 Geneva 23, Switzerland.}\\[5mm] 
$^3$~{\it Institut f\"ur Theoretische Teilchenphysik, Universit\"at Karlsruhe (TH),\\
          D-76128 Karlsruhe, Germany.}\\[25mm]

{\bf Abstract}\\[5mm]
\end{center} 
\setlength{\baselineskip}{0.2in} 

One of the most troublesome contributions to the NNLO QCD corrections to
${\bar B}\to X_s\gamma$ originates from three-loop matrix elements of
four-quark operators.  A part of this contribution that is proportional to the
QCD beta-function coefficient $\beta_0$ was found in 2003 as an expansion in
$m_c/m_b$.  In the present paper, we evaluate the asymptotic behaviour of the
complete contribution for $m_c \gg m_b/2$. The asymptotic form of the
$\beta_0$-part matches the small-$m_c$ expansion very well at the threshold
$m_c = m_b/2$.  For the remaining part, we perform an interpolation down to
the measured value of $m_c$, assuming that the $\beta_0$-part is a good
approximation at $m_c=0$. Combining our results with other contributions to
the NNLO QCD corrections, we find ${\cal B}({\bar B}\to X_s\gamma) = (3.15 \pm
0.23) \times 10^{-4}$ for $E_{\gamma} > 1.6\;$GeV in the ${\bar B}$-meson rest
frame. The indicated error has been obtained by adding in quadrature
the following uncertainties: non-perturbative (5\%), parametric (3\%),
higher-order 
perturbative (3\%), and the interpolation ambiguity (3\%).

\end{titlepage}

\newsection{Introduction \label{sec:intro}}

The decay ${\bar B}\to X_s\gamma$ is a well-known probe of new physics at the
electroweak scale. The current world average for its branching ratio with a
cut $E_{\gamma} > 1.6\;$GeV in the ${\bar B}$-meson rest frame reads~\cite{unknown:2006bi}
\be \label{eq:HFAG}
{\cal B}(\bar{B} \to X_s \gamma)_{\scs E_{\gamma} > 1.6\,{\rm GeV}}^{\scs\rm exp}
= \left(3.55\pm 0.24{\;}^{+0.09}_{-0.10}\pm0.03\right)\times 10^{-4}, 
\ee
where the first error is combined statistical and systematic. The
second one is due to the theory input on the shape function. The third
one is caused by the $b \to d\gamma$ contamination.

The total error in Eq.~(\ref{eq:HFAG}) amounts to around 7.4\%,
i.e. it is of the same size as the expected ${\cal O}(\alpha_s^2)$
corrections to the perturbative transition $b \to X_s^{\rm parton}
\gamma$. On the other hand, the relation
\be
\Gamma(\bar{B} \to X_s \gamma) ~\simeq~ \Gamma(b \to X_s^{\rm parton} \gamma)
\ee
holds up to non-perturbative corrections that turn out to be smaller (see
Section~\ref{sec:uncert}). \linebreak Consequently, evaluating the
Next-to-Next-to-Leading Order (NNLO) QCD corrections to \linebreak $b \to X_s^{\rm
  parton} \gamma$~ is of crucial importance for deriving constraints on new
physics from the measurements of~ ${\bar B}\to X_s\gamma$.

In the calculation of~ $b \to X_s^{\rm parton} \gamma$,~ resummation of large
logarithms~ $(\alpha_s \ln M_W^2/m_b^2)^n$~ is necessary at each order in
$\alpha_s$, which is most conveniently performed in the framework of an
effective theory that arises from the Standard Model (SM) after decoupling the
heavy electroweak bosons and the top quark. The explicit form of the relevant
effective Lagrangian is given in the next section. The Wilson coefficients
$C_i(\mu)$~play the role of coupling constants at the flavour-changing
vertices (operators) $Q_i$.

The perturbative calculations are performed in three steps: 
\begin{itemize}
\item[(i)] Matching:~ Evaluating $C_i(\mu_0)$ at the renormalization scale
  $\mu_0 \sim M_W,m_t$~ by requiring \linebreak equality of the SM and effective theory
  Green's functions at the leading order in \linebreak (external momenta)$/(M_W,m_t)$.
\item[(ii)] Mixing:~ Calculating the operator mixing under renormalization,
  deriving the effective theory Renormalization Group Equations (RGE) and
  evolving $C_i(\mu)$ from $\mu_0$ down to the low-energy scale $\mu_b \sim m_b$.
\item[(iii)] Matrix elements:~ Evaluating the on-shell~ $b \to X_s^{\rm parton}
  \gamma$~ amplitudes at~ $\mu_b \sim m_b$.
\end{itemize}
In the NNLO analysis of the considered decay, the four-quark operators
$Q_1,\ldots,Q_6$ and the dipole operators $Q_7$ and $Q_8$ must be matched at
the two- and three-loop level, respectively. Three-point amplitudes
with four-quark vertices need to be renormalized up to the four-loop
level, while ``only'' three-loop mixing is necessary in the remaining
cases. The matrix elements are needed up to two loops for the dipole
operators, and up to three loops for the four-quark operators.
  
The NNLO matching was calculated in Refs.~\cite{Bobeth:1999mk,Misiak:2004ew}.
The three-loop renormalization in the $\{Q_1,\ldots,Q_6\}$ and
$\{Q_7,Q_8\}$ sectors was found in Refs.~\cite{Gorbahn:2004my,Gorbahn:2005sa}.
The results from Ref.~\cite{Czakon:2006ss} on the four-loop
mixing of $Q_1,\ldots,Q_6$ into $Q_7$ will be used in our numerical
analysis.\footnote{
The small effect ($-0.35\%$ in the branching ratio)
    of the four-loop mixing~\cite{Czakon:2006ss} of $Q_1,\ldots,Q_6$
    into $Q_8$ is neglected here. It was not yet known in September 2006 when
    the current paper was being completed.}

As far as the matrix elements are concerned, contributions to the decay rate
that are proportional to $|C_7(\mu_b)|^2$ are completely known at the NNLO
thanks to the calculations in Refs.~\cite{Blokland:2005uk,Melnikov:2005bx}. 
These two-loop results have recently been confirmed by an independent
group~\cite{Asatrian:2006ph,Asatrian:2006sm}.  Two- and three-loop matrix
elements in the so-called large-$\beta_0$ approximation were found in
Ref.~\cite{Bieri:2003ue} as expansions in the quark mass ratio $m_c/m_b$.
Such expansions are adequate when $m_c < m_b/2$, which is satisfied by the
measured quark masses.  Finding all the remaining (``beyond-$\beta_0$'')
contributions to the matrix elements is a very difficult task because hundreds
of massive three-loop on-shell vertex diagrams need to be calculated.

In the present work, we evaluate the asymptotic form of the $m_c$-dependent
NNLO matrix elements in the limit $m_c \gg m_b/2$ using the same decoupling
technique as in our three-loop Wilson coefficient
calculation~\cite{Misiak:2004ew}.  We find that the asymptotic form of the
$\beta_0$-part matches the small-$m_c$ expansion very well at the $c\bar c$
production threshold $m_c = m_b/2$. The same is true for the Next-to-Leading
Order (NLO) matrix elements.  Motivated by this observation, we
interpolate the beyond-$\beta_0$ part to smaller values of $m_c$ assuming that
the $\beta_0$-part is a good approximation at $m_c=0$.  Combining our results
with other contributions to the NNLO QCD corrections, we find an estimate for
the branching ratio at ${\cal O}(\alpha_s^2)$.

Our paper is organized as follows. In Section~\ref{sec:Leff}, we introduce the
effective theory and collect the relevant formulae for the $\bar{B} \to X_s
\gamma$ branching ratio. The contributions that are known exactly in $m_c$ are
described in Section~\ref{sec:p21p32}. Expressions for the NNLO matrix
elements in the large-$\beta_0$ approximation and in the $m_c \gg m_b/2$ limit
are presented in Sections~\ref{sec:p22b0} and~\ref{sec:p22lim}, respectively.
Section~\ref{sec:interpol} is devoted to discussing the interpolation in
$m_c$. Section~\ref{sec:uncert} contains the analysis of uncertainties. We
conclude in Section~\ref{sec:conclusions}.  Our numerical input parameters are
collected in Appendix~A. Appendix~B contains a discussion of the 
$c\bar c$ production treatment in the interpolation.

\newsection{The effective theory \label{sec:Leff}}

Following Section~3 of Ref.~\cite{Gambino:2001ew}, the $\bar{B} \to X_s \gamma$
branching ratio can be expressed as follows:
\mathindent0cm
\be \label{main}
{\cal B}[\bar{B} \to X_s \gamma]_{\scs E_{\gamma} > E_0}
= {\cal B}[\bar{B} \to X_c e \bar{\nu}]_{\rm exp} 
\left| \f{ V^*_{ts} V_{tb}}{V_{cb}} \right|^2 
\f{6 \alpha_{\rm em}}{\pi\;C} 
\left[ P(E_0) + N(E_0) \right],
\ee
\mathindent1cm
where $\alpha_{\rm em} = \alpha_{\rm em}(0) \simeq 1/137.036$ and
$N(E_0)$ denotes the non-perturbative correction.\footnote{$\,$
  See Eqs.~(3.10) and (4.7) of Ref.~\cite{Gambino:2001ew}. The corrections
  found in Eqs.~(3.9) and (3.14) of Ref.~\cite{Bauer:1997fe}
  as well as Eq.~(28) of Ref.~\cite{Neubert:2004dd} 
  should be included in $N(E_0)$, too.}
The $m_c$-dependence of $\bar{B} \to X_c e \bar\nu$ is accounted for by
\be \label{phase1}
C = \left| \f{V_{ub}}{V_{cb}} \right|^2 
\f{\Gamma[\bar{B} \to X_c e \bar{\nu}]}{\Gamma[\bar{B} \to X_u e \bar{\nu}]},
\ee
with neglected spectator annihilation. $P(E_0)$ is given by the perturbative ratio
\be \label{pert.ratio}
\f{\Gamma[ b \to X_s \gamma]_{E_{\gamma} > E_0}}{
|V_{cb}/V_{ub}|^2 \; \Gamma[ b \to X_u e \bar{\nu}]} = 
\left| \f{ V^*_{ts} V_{tb}}{V_{cb}} \right|^2 
\f{6 \alpha_{\rm em}}{\pi} \; P(E_0).
\ee
Our goal is to calculate the NNLO QCD corrections to the quantity $P(E_0)$.
The denominator on the l.h.s. of Eq.~(\ref{pert.ratio}) is already known
  at the NNLO level from Refs.~\cite{vanRitbergen:1999gs,Steinhauser:1999bx}.

The relevant effective Lagrangian reads
\be 
{\cal L}_{\rm eff} = {\cal L}_{\scs {\rm QCD} \times {\rm QED}}(u,d,s,c,b) 
+ \f{4 G_F}{\sqrt{2}} \left[ V^*_{ts} V_{tb} \sum_{i=1}^{8} C_i Q_i
                            +  V^*_{us} V_{ub} \sum_{i=1}^{2} C^c_i (Q^u_i - Q_i) \right],
\label{Leff2}
\ee
where
\be \label{physical}
\begin{array}{rl}
Q^u_1 ~= & (\bar{s}_L \gamma_{\mu} T^a u_L) (\bar{u}_L \gamma^{\mu} T^a b_L), 
\vspace{0.2cm} \\
Q^u_2 ~= & (\bar{s}_L \gamma_{\mu}     u_L) (\bar{u}_L \gamma^{\mu}     b_L),
\vspace{0.2cm} \\
Q_1 ~= & (\bar{s}_L \gamma_{\mu} T^a c_L) (\bar{c}_L \gamma^{\mu} T^a b_L),
\vspace{0.2cm} \\
Q_2 ~= & (\bar{s}_L \gamma_{\mu}     c_L) (\bar{c}_L \gamma^{\mu}     b_L),
\vspace{0.2cm} \\
Q_3 ~= & (\bar{s}_L \gamma_{\mu}     b_L) \sum_q (\bar{q}\gamma^{\mu}     q),     
\vspace{0.2cm} \\
Q_4 ~= & (\bar{s}_L \gamma_{\mu} T^a b_L) \sum_q (\bar{q}\gamma^{\mu} T^a q),    
\vspace{0.2cm} \\
Q_5 ~= & (\bar{s}_L \gamma_{\mu_1}
                   \gamma_{\mu_2}
                   \gamma_{\mu_3}    b_L)\sum_q (\bar{q} \gamma^{\mu_1} 
                                                         \gamma^{\mu_2}
                                                         \gamma^{\mu_3}     q),     
\vspace{0.2cm} \\
Q_6 ~= & (\bar{s}_L \gamma_{\mu_1}
                   \gamma_{\mu_2}
                   \gamma_{\mu_3} T^a b_L)\sum_q (\bar{q} \gamma^{\mu_1} 
                                                          \gamma^{\mu_2}
                                                          \gamma^{\mu_3} T^a q),
\vspace{0.2cm} \\
Q_7  ~= &  \f{e}{16\pi^2} m_b (\bar{s}_L \sigma^{\mu \nu}     b_R) F_{\mu \nu},
\vspace{0.2cm} \\
Q_8  ~= &  \f{g}{16\pi^2} m_b (\bar{s}_L \sigma^{\mu \nu} T^a b_R) G_{\mu \nu}^a.
\end{array}
\ee

The last term in the square bracket of Eq.~(\ref{Leff2}) gives no
contribution at the Leading Order (LO) and only a small contribution at
the NLO (around $+1\%$ in the branching ratio --- see Eq.~(3.7) of
Ref.~\cite{Gambino:2001ew}). Consequently, we shall neglect its effect on the
NNLO QCD correction and omit terms proportional to $V_{ub}$ in the analytical
formulae below. However, our numerical results will include the $V_{ub}$ terms
at the NLO. The same refers to the electroweak corrections that amount to
around $-3.7\%$ in $P(E_0)$~\cite{Gambino:2001ew,Gambino:2001au}.

The quantity $P(E_0)$ depends quadratically on the Wilson coefficients\footnote{$\,$
In Eq.~(30) of Ref.~\cite{Blokland:2005uk}, $K_{ij}$ was denoted by $\widetilde{G}_{ij}/G_u$.}
\be \label{Kij1}
P(E_0) = \sum_{i,j=1}^{8} C_i^{\rm eff}(\mu_b)\; C_j^{\rm eff}(\mu_b)\; K_{ij}(E_0,\mu_b),
\ee
where the ``effective coefficients'' are defined by 
\be
C_i^{\rm eff}(\mu) = \left\{ \begin{array}{ll}
C_i(\mu), & \mbox{ for $i = 1, ..., 6$,} \\[1mm] 
C_7(\mu) + \sum_{j=1}^6 y_j C_j(\mu), & \mbox{ for $i = 7$,} \\[1mm]
C_8(\mu) + \sum_{j=1}^6 z_j C_j(\mu), & \mbox{ for $i = 8$.}
\end{array} \right.
\ee
The numbers $y_j$ and $z_j$ are defined so that the leading-order $b \to s
\gamma$ and $b \to sg$ matrix elements of the effective Hamiltonian are
proportional to the leading-order terms in $C_7^{\rm eff}$ and
$C_8^{\rm eff}$, respectively~\cite{Buras:1994xp}. This means, in
particular, that $K_{ij} = \delta_{i7}\delta_{j7} + {\cal O}(\alpha_s)$. In
the $\overline{\rm MS}$ scheme with fully anticommuting $\gamma_5$,~ $\vec{y} = (0, 0,
-\f{1}{3}, -\f{4}{9}, -\f{20}{3}, -\f{80}{9})$ and $\vec{z} = (0, 0, 1, -\f{1}{6},
20, -\f{10}{3})$~\cite{Chetyrkin:1996vx}.

In Eq.~(\ref{Kij1}), we have assumed that all the Wilson coefficients are real, as it
is the case in the SM. Consequently, $K_{ij}$ is a real symmetric matrix.

Once the $\overline{\rm MS}$-renormalized\footnote{$\,$
The evanescent operators are as in Eqs.~(23)--(25) of Ref.~\cite{Gorbahn:2004my}.}
coefficients $C_i^{\rm eff}(\mu)$ are perturbatively expanded
\be
C_i^{\rm eff}(\mu) = C^{(0)\rm eff}_i(\mu) + \al(\mu) C^{(1)\rm eff}_i(\mu) + \al^2(\mu)
C^{(2)\rm eff}_i(\mu) + {\cal O}\left(\al^3(\mu)\right), \nnb\\ 
\ee
where
\be 
\al(\mu_b) \equiv \f{\alpha_s^{(5)}(\mu_b)}{4\pi}, 
\ee
the expression for $P(E_0)$ can be cast in the following form:
\bea
P(E_0) &=& P^{(0)}(\mu_b) 
+ \al(\mu_b) \left[ P_1^{(1)}(\mu_b) + P_2^{(1)}(E_0,\mu_b) \right] \nnb\\ 
&+& \al^2(\mu_b) \left[ P_1^{(2)}(\mu_b) + P_2^{(2)}(E_0,\mu_b) + P_3^{(2)}(E_0,\mu_b) \right]
+ {\cal O}\left(\al^3(\mu_b)\right). \label{Pexp}
\eea
Here, $P^{(0)}$ and  $P_1^{(k)}$ originate from the tree-level matrix element
of $Q_7$
\bea
P^{(0)}(\mu_b)   &=&   \left( C_7^{(0)\rm eff}(\mu_b)\right)^2, \nnb\\
P_1^{(1)}(\mu_b) &=& 2 C_7^{(0)\rm eff}(\mu_b) C_7^{(1)\rm eff}(\mu_b), \nnb\\
P_1^{(2)}(\mu_b) &=&   \left( C_7^{(1)\rm eff}(\mu_b)\right)^2 + 2 C_7^{(0)\rm eff}(\mu_b) C_7^{(2)\rm eff}(\mu_b),
\eea
while $P_2^{(k)}$ depend only on the LO Wilson coefficients 
$C_i^{(0)\rm eff}$. The NNLO correction $P_3^{(2)}$ is defined by requiring
that it is proportional to products of the LO and NLO Wilson coefficients only
($C_i^{(0)\rm eff} C_j^{(1)\rm eff}$).

\newsection{The corrections $P_2^{(1)}$ and $P_3^{(2)}$ \label{sec:p21p32}}

The corrections $P_2^{(1)}$ and $P_3^{(2)}$ are known exactly in $m_c$.  In
order to describe their content, we expand $K_{ij}(E_0,\mu_b)$ in $\al(\mu_b)$
\be
K_{ij} = \delta_{i7}\delta_{j7} + \al(\mu_b) K_{ij}^{(1)} + \al^2(\mu_b)
K_{ij}^{(2)} + {\cal O}\left(\al^3(\mu_b)\right).
\ee  
The coefficients $K_{ij}^{(1)}$ are easily derived from the known NLO results
\bea \label{ki7}
K_{i7}^{(1)} &=&  {\rm Re}\,r_i^{(1)} - \f{1}{2} \gamma^{(0)\rm eff}_{i7} L_b + 2 \phi_{i7}^{(1)}(\delta),
\hspace{1cm} {\rm for~} i \leq 6,\\
K_{77}^{(1)} &=& -\f{182}{9} + \f{8}{9}\pi^2 - \gamma^{(0)\rm eff}_{77} L_b + 4\,\phi_{77}^{(1)}(\delta), \\
K_{78}^{(1)} &=&  \f{44}{9} - \f{8}{27}\pi^2 -\f{1}{2} \gamma^{(0)\rm eff}_{87} L_b  + 2\,\phi_{78}^{(1)}(\delta),\\
K_{ij}^{(1)} &=&  2 (1 + \delta_{ij}) \phi_{ij}^{(1)}(\delta),
\hspace{34mm} {\rm for~} i,j \neq 7, \label{kij}
\eea
where 
\be
L_b = \ln \left( \f{\mu_b}{m_b^{1S}} \right)^2. 
\ee
The matrix $\hat{\gamma}^{(0)\rm eff}$ and the quantities $r_i^{(1)}$ as
functions of
\be
z= \left(\f{m_c(\mu_c)}{m_b^{1S}}\right)^2
\ee
can be found respectively in Eqs.~(6.3) and (3.1) of Ref.~\cite{Buras:2002tp}.
The bottom mass is renormalized in the $1S$ scheme \cite{Hoang:1999ye}
  throughout the paper. The charm mass $\overline{\rm MS}$
renormalization scale $\mu_c$ is chosen to be independent from $\mu_b$. For
future convenience, we quote $r_{1,2}^{(1)}$:
\be
r_2^{(1)}(z) = -6\,r_1^{(1)}(z) = -\f{1666}{243} + 2[a(z)+b(z)] - \f{80}{81} i \pi.
\ee
The exact expressions for $a(z)$ and $b(z)$ in terms of Feynman parameter
integrals can be found in Eqs.~(3.3) and (3.4) of Ref.~\cite{Buras:2002tp}.
Their small-$m_c$ expansions up to ${\cal O}(z^4)$ read~\cite{Greub:1996tg,Buras:2001mq}
\mathindent0cm
\bea
a(z)  &=& \f{16}{9} \left\{ 
\left[ \f{5}{2} -\f{\pi^2}{3} -3 \zeta(3) + \left( \f{5}{2} - \f{3\pi^2}{4} \right) L_z
+ \f{1}{4} L_z^2 + \f{1}{12} L_z^3 \right] z + \left(\f{7}{4} +\f{2\pi^2}{3} -\f{\pi^2}{2} L_z 
\right.\right.\nnb\\[2mm]&-&\left.\left.
\f{1}{4} L_z^2 +\f{1}{12} L_z^3 \right) z^2 
+ \left[ -\f{7}{6} -\f{\pi^2}{4} + 2L_z - \f{3}{4}L_z^2 \right] z^3
+ \left( \f{457}{216} - \f{5\pi^2}{18} -\f{1}{72} L_z -\f{5}{6} L_z^2 \right) z^4 
\right.\nnb\\[2mm]&+&\left.\hspace{-1mm}
 i \pi \left[ \left( 4 -\f{\pi^2}{3} + L_z + L_z^2 \right) \f{z}{2} 
+ \left( \f{1}{2} -\f{\pi^2}{6} - L_z\! + \f{1}{2} L_z^2 \right) z^2 + z^3 
+ \f{5}{9} z^4  \right]\right\} + {\cal O}(z^5 L_z^2),\\[2mm]
b(z) &=& -\f{8}{9} \left\{
\left( -3 +\f{\pi^2}{6} - L_z \right) z - \f{2\pi^2}{3} z^{3/2}
+ \left( \f{1}{2} + \pi^2 - 2 L_z - \f{1}{2} L_z^2 \right) z^2 
\right.\nnb\\[2mm]&+&\left.
\left(-\f{25}{12} -\f{1}{9} \pi^2 - \f{19}{18} L_z + 2 L_z^2 \right) z^3 
+ \left( -\f{1376}{225} + \f{137}{30} L_z + 2 L_z^2 + \f{2\pi^2}{3} \right) z^4
\right.\nnb\\[2mm]&+&\left.
i \pi \left[ -z + (1-2 L_z) z^2 
+ \left(-\f{10}{9} + \f{4}{3} L_z \right) z^3 + z^4\right]\right\}
+ {\cal O}(z^5 L_z^2),
\eea
\mathindent1cm
where 
\be
L_z = \ln z.
\ee
The functions $\phi_{ij}^{(1)}(\delta \equiv 1 - 2E_0/m_b^{1S})$ with $i,j \in
\{1,2,7,8\}$ can be found in Appendix~E of Ref.~\cite{Gambino:2001ew}. The
remaining ones (that affect $P(1.6\,{\rm GeV})$ by $\sim 0.1\%$ only) can be
read out from the results of Ref.~\cite{Pott:1995if}. In particular,
\bea
\phi_{47}^{(1)}(\delta) &=& -\f{1}{54} \delta \left( 1 - \delta + \f{1}{3} \delta^2 \right)~
+~ \f{1}{12}~ \lim_{m_c \to m_b} \phi_{27}^{(1)}(\delta),\\
\phi_{48}^{(1)}(\delta) &=& -\f{1}{3} \phi_{47}^{(1)}(\delta).
\eea

Once all the ingredients of $K_{ij}^{(1)}$ have been specified, $P_2^{(1)}$
and $P_3^{(2)}$ are evaluated by simple substitutions to Eq.~(\ref{Kij1})
\bea
P_2^{(1)} &=& \sum_{i,j=1}^{8} C_i^{(0)\rm eff}\; C_j^{(0)\rm eff} \; K_{ij}^{(1)},\label{p21}\\[2mm] 
P_3^{(2)} &=& 2 \sum_{i,j=1}^{8} C_i^{(0)\rm eff}\; C_j^{(1)\rm eff} \; K_{ij}^{(1)}.\label{p32}
\eea

\newsection{The $\beta_0$-part of $P_2^{(2)}$ \label{sec:p22b0}}

The only NNLO correction to $P(E_0)$ in Eq.~(\ref{Pexp}) that has not yet been
given is $P_2^{(2)}$. For this contribution, we shall neglect the tiny LO
Wilson coefficients of $Q_3,\ldots,Q_6$. The NLO matrix elements of these
operators affect the branching ratio by only around 1\%~\cite{Buras:2002tp}.
Thus, neglecting the corresponding NNLO ones has practically no influence on
the final accuracy.

Let us split $K_{ij}^{(2)}$ into the $\beta_0$-parts $K_{ij}^{(2)\beta_0}$ and the
remaining parts $K_{ij}^{(2)\rm rem}$
\be \label{ksplit}
K_{ij}^{(2)} ~=~ A_{ij}\, n_f + B_{ij} ~=~ K_{ij}^{(2)\beta_0} + K_{ij}^{(2)\rm rem}, 
\ee
where $n_f$ stands for the number of massless flavours in the effective
theory, and
\be
K_{ij}^{(2)\beta_0} \equiv -\f{3}{2}\beta_0 A_{ij} = -\f{3}{2} \left(11 - \f{2}{3} n_f\right) A_{ij},
\hspace{2cm}
K_{ij}^{(2)\rm rem} \equiv \f{33}{2} A_{ij} + B_{ij}.
\ee
Following Ref.~\cite{Bieri:2003ue}, we shall take $n_f=5$. Effects
  related to the absence of real $c\bar c$ production in $b \to X_s^{\rm
    parton} \gamma$ and to non-zero masses in quark loops on gluon propagators
  are relegated to $K_{ij}^{(2)\rm rem}$. Thus, the only $m_c$-dependent
contributions to $K_{ij}^{(2)\beta_0}$ originate from charm loops containing
the four-quark vertices $Q_1$ and $Q_2$.

The explicit $K_{ij}^{(2)\beta_0}$ that we derive from the results of
Refs.~\cite{Blokland:2005uk,Melnikov:2005bx,Bieri:2003ue,Buras:2002tp,vanRitbergen:1999gs}
read
%
\bea \label{k27b0}
K_{27}^{(2)\beta_0} &=& \beta_0\, 
{\rm Re}\left\{ -\f{3}{2} r_2^{(2)}(z) +2\left[a(z)+b(z)-\f{290}{81}\right] L_b 
- \f{100}{81} L_b^2 \right\} + 2 \phi^{(2)\beta_0}_{27}(\delta),\\[1mm]
K_{17}^{(2)\beta_0} &=& -\f{1}{6} K_{27}^{(2)\beta_0},\\[1mm]
K_{77}^{(2)\beta_0} &=& \beta_0 \left\{ -\f{3803}{54} -\f{46}{27} \pi^2 +\f{80}{3} \zeta(3) 
+ \left( \f{8}{9} \pi^2 - \f{98}{3} \right) L_b -\f{16}{3} L_b^2 \right\} + 4\phi_{77}^{(2)\beta_0}(\delta),\\[1mm]
K_{78}^{(2)\beta_0} &=& \beta_0 \left\{ \f{1256}{81} -\f{64}{81} \pi^2 - \f{32}{9} \zeta(3)
+ \left( \f{188}{27} -\f{8}{27} \pi^2 \right) L_b + \f{8}{9} L_b^2  \right\} + 2\phi_{78}^{(2)\beta_0}(\delta),
\label{k78b0}\\[1mm]
K_{ij}^{(2)\beta_0} &=&
2(1+\delta_{ij})\phi^{(2)\beta_0}_{ij}(\delta),\hspace{1cm} {\rm for}~i,j \neq 7.
\eea
The small-$m_c$ expansion of ${\rm Re}\,r_2^{(2)}(z)$
up to ${\cal O}(z^4)$ was calculated by Bieri {\it et al.}~\cite{Bieri:2003ue} 
\mathindent0cm
\bea
{\rm Re}\,r_2^{(2)}(z) &=& \f{67454}{6561} - \f{124\pi^2}{729} -
  \f{4}{1215} \left(11280 - 1520\pi^2 - 171\pi^4 - 5760 \zeta(3) 
   + 6840L_z 
\right.\nnb\\[1mm]&-&\left.
1440\pi^2L_z - 2520\zeta(3)L_z + 120L_z^2 + 100 L_z^3 - 30 L_z^4\right) z 
\nnb\\&-&
\f{64\pi^2}{243} \left( 43 - 12 \ln 2 -3L_z\right) z^{3/2} 
- \f{2}{1215} \left(11475 - 380\pi^2 + 96\pi^4 +7200 \zeta(3) 
\right.\nnb\\&-&\left.
1110L_z - 1560\pi^2L_z + 1440
  \zeta(3)L_z +990L_z^2 +260L_z^3 -60L_z^4\right)z^2 
\nnb\\&+&
\f{2240\pi^2}{243} z^{5/2} - \f{2}{2187} \left(62471 -2424\pi^2
  - 33264\zeta(3) - 19494L_z - 504\pi^2L_z 
\right.\nnb\\&-&\left.
5184L_z^2 +2160L_z^3\right) z^3 - \f{2464}{6075} \pi^2 z^{7/2}
+ \left( - \f{15103841}{546750} + \f{7912}{3645} \pi^2 + \f{2368}{81} \zeta(3) 
\right.\nnb\\&+&\left.
\f{147038}{6075} L_z + \f{352}{243} \pi^2 L_z 
+ \f{88}{243} L_z^2 - \f{512}{243} L_z^3 \right) z^4 
+ {\cal O}(z^{9/2} L_z^4). \label{rer2}
\eea
\mathindent1cm

The function $\phi_{77}^{(2)\beta_0}(\delta)$ reads
\be
\phi_{77}^{(2)\beta_0}(\delta) = \beta_0 \left[ \phi_{77}^{(1)}(\delta) L_b 
+ 4 \int_0^{1-\delta} {\mathrm d}x\; F^{(2,nf)} \right],
\ee
where $F^{(2,nf)}$ as a function of $x = 2 E_{\gamma}/m_b$ 
is given in Eq.~(9) of Ref.~\cite{Melnikov:2005bx}.\footnote{$\,$
The original calculation of $F^{(2,nf)}$ and several other
contributions to the photon spectrum was performed in Ref.~\cite{Ligeti:1999ea}.}

The remaining functions $\phi_{ij}^{(2)\beta_0}(\delta)$ will
be neglected in our numerical analysis. It should not cause any
significant uncertainty for $E_0 = 1.6\,$GeV. For this particular cut,
the NLO functions $\phi_{ij}^{(1)}(\delta)$ affect the branching ratio
by around $-4\%$ only, which is partly due to a certain convention in
their definitions ($\phi_{ij}^{(1)}(0)=0$ for $(ij)\neq(77)$, and
$\phi_{77}^{(1)}(1)=0$).  An analogous convention is used for
$\phi_{ij}^{(2)\beta_0}(\delta)$.  The known
$\phi_{77}^{(2)\beta_0}(\delta)$ affects the branching ratio by around
$-0.4\%$ only.  If the NLO pattern is repeated, an effect of
similar magnitude is expected from the other
$\phi_{ij}^{(2)\beta_0}(\delta)$.

As in Eq.~(\ref{ksplit}), we can split $P_2^{(2)} = P_2^{(2)\beta_0} +
P_2^{(2)\rm rem}$ and express $P_2^{(2)\beta_0}$ in terms of
$K_{ij}^{(2)\beta_0}$, by analogy to Eq.~(\ref{p21}) 
\be
P_2^{(2)\beta_0} \simeq \sum_{i,j=1,2,7,8} 
C_i^{(0)\rm eff}\; C_j^{(0)\rm eff}\; K_{ij}^{(2)\beta_0}. \label{p2beta0}
\ee
The ``$\simeq$'' sign is used above only because we skip $i,j=3,4,5,6$ 
in the sum.

\newsection{The full correction $P_2^{(2)}$ in the limit $m_c \gg m_b/2$ \label{sec:p22lim}}

The present section contains the main new result of our paper, namely the
asymptotic form of $P_2^{(2)}$ in the limit $m_c \gg m_b/2$. It has been
evaluated by performing a formal three-loop decoupling of the charm
quark in the effective theory, using the method that has been previously
applied by us to the calculation of the three-loop matching at the
electroweak scale~\cite{Misiak:2004ew,Steinhauser:2000ry}.  Once the charm
decoupling scale is set equal to $\mu_b$, one recovers the asymptotic form of
the matrix elements in the large $m_c$ limit. Details of this calculation
will be presented elsewhere \cite{decoupling.details}.

The $m_c$-dependence of $P(E_0)$ at the NLO is dominated by ${\rm
  Re}[a(z)+b(z)]$.  At the two-loop level, we find the following asymptotic
form of the functions $a(z)$ and $b(z)$
\bea 
{\rm Re}\,a(z) &=& \f{4}{3} L_z + \f{34}{9} 
+ \f{1}{z} \left( \f{5}{27} L_z + \f{101}{486} \right) 
+ \f{1}{z^2} \left( \f{1}{15} L_z + \f{1393}{24300} \right) 
+ {\cal O}\left(\f{1}{z^3}\right), \label{aasymp}\\[2mm]
{\rm Re}\,b(z) &=& -\f{4}{81} L_z + \f{8}{81} 
- \f{1}{z} \left( \f{2}{45} L_z + \f{76}{2025} \right) 
- \f{1}{z^2} \left( \f{4}{189} L_z + \f{487}{33075} \right) 
+ {\cal O}\left(\f{1}{z^3}\right). \label{basymp}
\eea
${\rm Im}\,a(z)=\f{4}{9}i\pi$~ and~ ${\rm Im}\,b(z)=\f{4}{81}i\pi$~ are
exactly constant for~ $z > \f{1}{4}$.

For the real part of the three-loop function 
$r_2^{(2)}(z)$ introduced in Eq.~(\ref{k27b0}), we find
\bea 
{\rm Re}\,r_2^{(2)}(z) &=& \f{8}{9}L_z^2 + \f{112}{243} L_z + \f{27650}{6561} 
+ \f{1}{z} \left( \f{38}{405} L_z^2 - \f{572}{18225} L_z + \f{10427}{30375} - \f{8}{135} \pi^2 \right)\nnb\\[2mm]
&+& \f{1}{z^2} \left( \f{86}{2835} L_z^2 - \f{1628}{893025} L_z + \f{19899293}{125023500} - \f{8}{405} \pi^2 \right)
+ {\cal O}\left(\f{1}{z^3}\right).
\eea
The small-$m_c$ expansion of this function has been given in Eq.~(\ref{rer2}).

\begin{figure}[t]
\begin{center}
\includegraphics[width=8cm,angle=0]{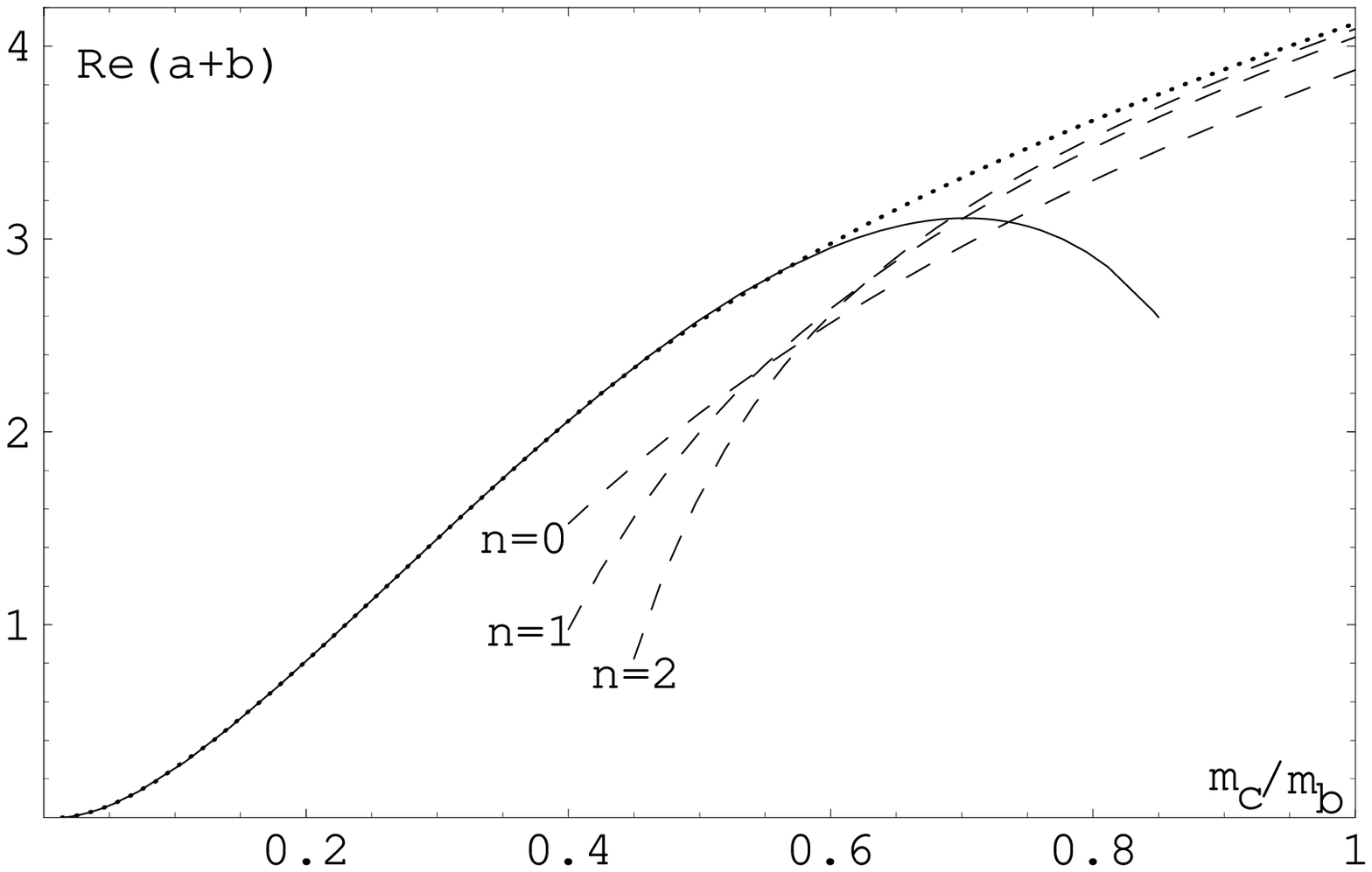}
\hspace{5mm}
\includegraphics[width=8cm,angle=0]{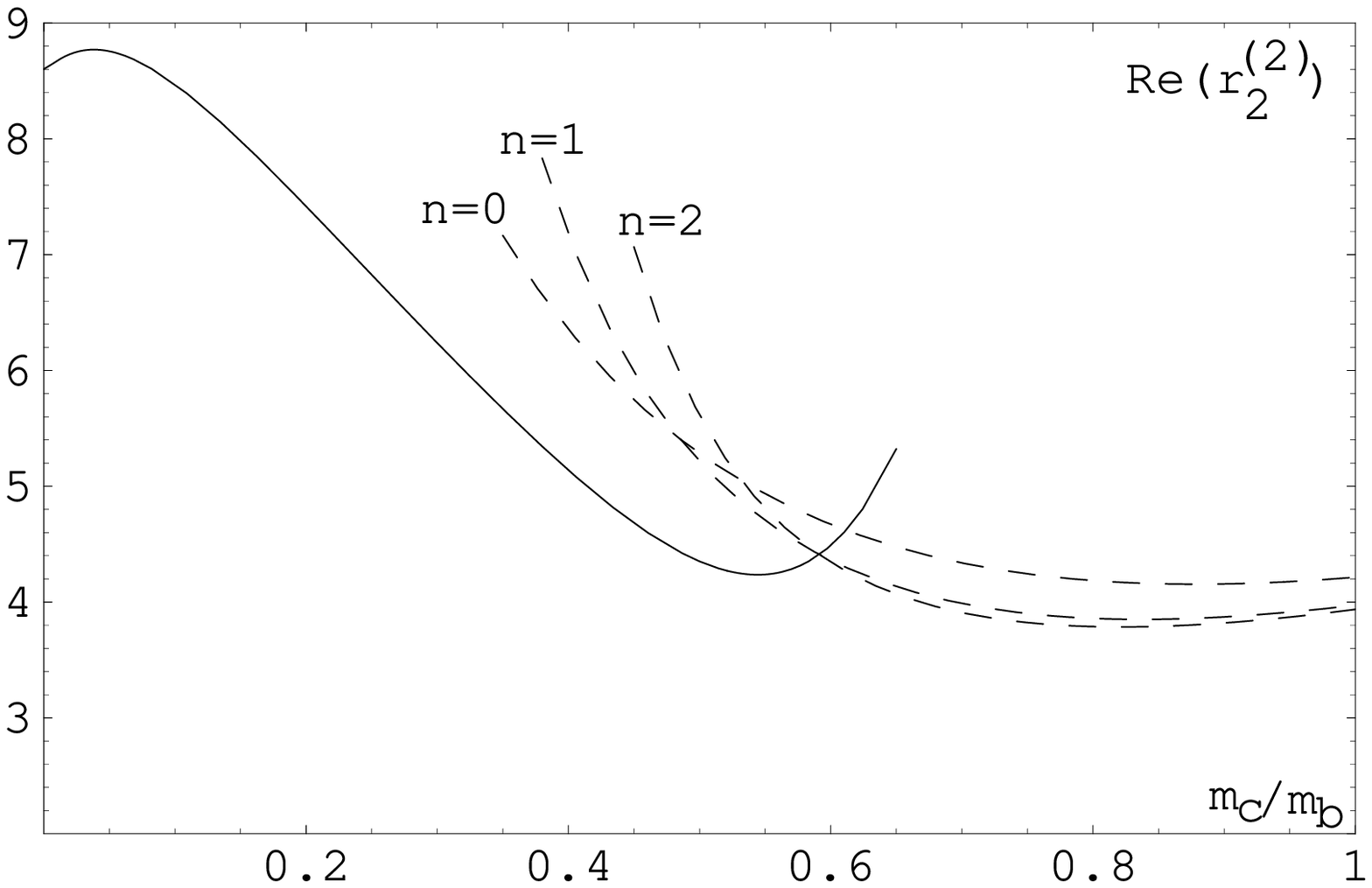} 
\caption{\sf ${\rm Re}(a+b)$ (left plot) and ${\rm Re}\,r_2^{(2)}$ (right
  plot) as functions of $m_c/m_b = \sqrt{z}$. See the text for 
  explanation of the curves. \label{fig:abr2}}
\end{center}
\end{figure}

The dotted line in the left plot of Fig.~\ref{fig:abr2} corresponds to the
exact expression for ${\rm Re}(a+b)$. The solid line presents its small-$m_c$
expansion up to ${\cal O}(z^4)$. The dashed lines are found from the
large-$m_c$ expansion including terms up to ${\cal O}(z^{-n})$ with $n=0,1,2$,
which is indicated by labels at the curves.  The solid and dashed lines in the
right plot of Fig.~\ref{fig:abr2} present the same expansions of ${\rm
  Re}\,r_2^{(2)}$. No exact expression is known in this case.

The plots in Fig.~\ref{fig:abr2} clearly demonstrate that a combination of the
small-$m_c$ expansion for $m_c < m_b/2$ and the large-$m_c$ expansion for $m_c
> m_b/2$ (even in the $n=0$ case) leads to a reasonable approximation to
${\rm Re}(a+b)$ and ${\rm Re}\,r_2^{(2)}$ for any $m_c$. Moreover, no large
$c\bar{c}$ threshold effects are seen at $m_c = m_b/2$.  These observations
motivate us to calculate the $n=0$ term in the large-$m_c$ expansion of
$P_2^{(2)\rm rem}$ and use it in Section~\ref{sec:interpol} to estimate this
quantity for $m_c < m_b/2$.

The expressions that we have found for the leading terms in the large-$m_c$
expansion of $K_{ij}^{(2)\rm rem}$ are presented below. The necessary leading
terms of $K_{ij}^{(1)}$ are easily derived from Eqs.~(\ref{aasymp}) and
(\ref{basymp}), taking into account that only $\phi_{ij}^{(1)}$ with
$i,j > 2$ do not vanish at large $z$, and that $\phi_{ij}^{(1)}$ are $z$-independent
for $i,j=4,7,8$.
\mathindent0cm
\bea 
K_{22}^{(2)\rm rem} &=&
36\,K_{11}^{(2)\rm rem} + {\cal O}\left(\f{1}{z}\right) ~=~
-6\,K_{12}^{(2)\rm rem} + {\cal O}\left(\f{1}{z}\right) ~=~
\left(K_{27}^{(1)}\right)^2 + {\cal O}\left(\f{1}{z}\right) \nnb\\[2mm]
&=& \left[ \f{218}{243} - \f{208}{81} L_D\right]^2 
+ {\cal O}\left(\f{1}{z}\right), \label{k22rem}\\[2mm]
K_{27}^{(2)\rm rem} &=& K_{27}^{(1)} K_{77}^{(1)} 
+ \left( \f{127}{324} - \f{35}{27} L_D \right) K_{78}^{(1)} 
+ \f{2}{3} (1 - L_D ) K_{47}^{(1)\rm rem}
\nnb\\[2mm] &-&
\f{4736}{729} L_D^2 + \f{1150}{729} L_D - \f{1617980}{19683} + \f{20060}{243} \zeta(3) 
+ \f{1664}{81} L_c + {\cal O}\left(\f{1}{z}\right),\\[2mm]
K_{28}^{(2)\rm rem} &=& K_{27}^{(1)} K_{78}^{(1)} 
+ \left( \f{127}{324} - \f{35}{27} L_D \right) K_{88}^{(1)} 
+ \f{2}{3} (1 - L_D ) K_{48}^{(1)}  + {\cal O}\left(\f{1}{z}\right),\\[2mm]
K_{17}^{(2)\rm rem} &=&\!\! -\f{1}{6} K_{27}^{(2)\rm rem} + \left(\f{5}{16}-\f{3}{4}L_D\right)K_{78}^{(1)}
-\f{1237}{729} +\f{232}{27}\zeta(3) +\f{70}{27}L_D^2 -\f{20}{27} L_D
+ {\cal O}\left(\f{1}{z}\right)\!,\\[2mm]
K_{18}^{(2)\rm rem} &=&\!\! -\f{1}{6} K_{28}^{(2)\rm rem} + \left(\f{5}{16}-\f{3}{4}L_D\right)K_{88}^{(1)}
+ {\cal O}\left(\f{1}{z}\right),
\eea
\ \\[-1cm]
\bea
K_{77}^{(2)\rm rem} &=& \left( K_{77}^{(1)} - 4 \phi_{77}^{(1)}(\delta)
+\f{2}{3} L_z \right) K_{77}^{(1)} -\f{32}{9} L_D^2 + \f{224}{27} L_D 
-\f{628487}{729} - \f{628}{405} \pi^4 
\nnb\\&+& 
\f{31823}{729} \pi^2 + \f{428}{27} \pi^2 \ln 2 + \f{26590}{81} \zeta(3) 
- \f{160}{3} L_b^2 - \f{2720}{9} L_b + \f{256}{27} \pi^2 L_b
\nnb\\&+& \f{512}{27}\pi\alpha_\Upsilon\; +\; 4 \phi_{77}^{(2)\rm rem}(\delta)\;
+ {\cal O}\left(\f{1}{z}\right), \label{k77rem}\\[2mm]
K_{78}^{(2)\rm rem} &=&  \left(-\f{50}{3} + \f{8}{3} \pi^2 - \f{2}{3} L_D \right) K_{78}^{(1)} 
+ \f{16}{27} L_D^2 -\f{112}{81} L_D + \f{364}{243} + X_{78}^{(2)\rm rem} 
+ {\cal O}\left(\f{1}{z}\right), \label{k78rem}\\[2mm]
K_{88}^{(2)\rm rem} &=& \left(-\f{50}{3} + \f{8}{3} \pi^2 - \f{2}{3} L_D \right) K_{88}^{(1)} 
+ X_{88}^{(2)\rm rem} + {\cal O}\left(\f{1}{z}\right), \label{k88rem}
\eea
\mathindent1cm
where~ 
\bea
K_{47}^{(1)\rm rem} &=& K_{47}^{(1)} - \beta_0 \left( \f{26}{81} - \f{4}{27} L_b \right),\\[2mm]
L_c &=& \ln \left( \f{\mu_c}{m_c(\mu_c)}\right)^2, 
\eea
and the ``decoupling logarithm''
\be
L_D \equiv L_b - L_z = \ln \left( \f{\mu_b}{m_c(\mu_c)}\right)^2.
\ee
The function $\phi_{77}^{(2)\rm rem}(\delta)$ reads
\mathindent0cm
\bea
\phi_{77}^{(2)\rm rem}(\delta) &=& -4\int_0^{1-\delta} {\mathrm d}x\; 
\left[ \f{16}{9} F^{(2,a)} + 4 F^{(2,na)} +\f{29}{3} F^{(2,nf)} \right]~
\nnb\\[3mm]&-& 
\f{8\pi\,\alpha_\Upsilon }{27\,\delta} 
\left[ 2 \delta \ln^2 \delta 
+ \left(4 +\! 7 \delta -\! 2 \delta^2 +\! \delta^3\right)\ln\delta
+ 7 - \f{8}{3}\delta - 7\delta^2 + 4\delta^3 - \f{4}{3}\delta^4 \right], \label{phi77rem}
\eea
\mathindent1cm
where $F^{(2,a)}$, $F^{(2,na)}$ and $F^{(2,nf)}$ as functions of
$x = 2 E_{\gamma}/m_b$ are given in Eqs.~(7)--(9) of Ref.~\cite{Melnikov:2005bx}.
The terms proportional to $\alpha_{\Upsilon} \equiv \alpha_s^{(4)}(\mu=m_b^{1S})$
occur in Eqs.~(\ref{k77rem}) and (\ref{phi77rem}) because the so-called Upsilon 
expansion prescription~\cite{Hoang:1998hm} is followed here, which means
that $m_b^{\rm pole}$ is expressed in terms of $m_b^{1S}$, and
$\alpha_{\Upsilon}$ in the ratio
\be
\f{m_b^{1S}}{m_b^{\rm pole}} = 1 - \f{8 \pi}{9} \al(\mu_b) \alpha_\Upsilon
+ \ldots
\ee
is treated as independent from $\al(\mu_b)$, i.e. it is not included in the
order-counting when the ${\cal O}(\al^3(\mu_b))$ terms are neglected. A reader
who wants to bypass this prescription and use another kinetic scheme for $m_b$
should set $\alpha_{\Upsilon}$ to zero in Eqs.~(\ref{k77rem}) and
(\ref{phi77rem}) and at the same time replace $m_b^{1S}$ by $m_b^{\rm pole}$.
Next, the latter mass should be perturbatively re-expressed in terms of the
chosen $m_b^{\rm kinetic}$ to get rid of the renormalon ambiguity. We have
verified that the $\sim 0.4\%$ effect of the $\alpha_{\Upsilon}$-terms
on the branching ratio can be reproduced by using 
$m_b^{\rm pole} = 4.74\;{\rm GeV}$ at the considered order.

The $m_c$-independent quantities $X_{78}^{(2)\rm rem}$ and 
$X_{88}^{(2)\rm rem}$ in Eqs.~(\ref{k78rem}) and (\ref{k88rem}) stand for the
unknown non-$\beta_0$ contributions from the two-loop matrix element of $Q_8$
in the theory with decoupled charm (together with the corresponding
bremsstrahlung).  They will be set to zero in the numerical analysis.
Neglecting them can be justified by arguing that the contribution of
$K_{78}^{(2)}$ to $P(E_0)$ is suppressed relative to that of $K_{77}^{(2)}$ by
$|Q_d C_8^{\rm eff}/C_7^{\rm eff}|
\simeq \f{1}{6}$. 
In effect, $K_{78}^{(2)\beta_0}$ in Eq.~(\ref{k78b0}) affects $P(E_0)$ by around
0.1\% only. The suppression factor gets squared for $K_{88}^{(2)}$. 
Thus, neglecting $X_{78}^{(2)\rm rem}$ and $X_{88}^{(2)\rm rem}$ is not
expected to cause any significant uncertainty.

Using the results of Refs.~\cite{Blokland:2005uk,Melnikov:2005bx}, one easily
derives an expression for the $m_c\to0$ limit of
  $\widetilde{K}_{77}^{(2)\rm rem}$ that differs from $K_{77}^{(2)\rm rem}$ by
  inclusion of the real $c\bar c$ production contributions (see Appendix B)
\bea
\widetilde{K}_{77}^{(2)\rm rem}(z=0) &=&
\left( K_{77}^{(1)} - 4 \phi_{77}^{(1)}(\delta) +\f{2}{3} L_b \right) K_{77}^{(1)} 
- \f{587708}{729} + \f{32651}{729} \pi^2 - \f{628}{405} \pi^4 
\nnb\\[1mm]&+& 
\f{428}{27} \pi^2 \ln 2 + \f{25150}{81} \zeta(3) 
- \f{448}{9} L_b^2 + \left( \f{80}{9}\pi^2 - \f{2524}{9} \right) L_b
\nnb\\[1mm]&+& 
\f{512}{27}\pi\alpha_\Upsilon\; +\; 4 \phi_{77}^{(2)\rm rem}(\delta)\;
- \f{8 \phi_{77}^{(2)\beta_0}}{3 \beta_0}. \label{k772remmc0}
\eea
We are going to use this formula for testing the $m_c$-interpolation
prescription in the next section.

By analogy to Eqs.~(\ref{p21}) and (\ref{p2beta0}), we can write $P_2^{(2)\rm rem}$ as
\be
P_2^{(2)\rm rem} \simeq \sum_{i,j=1,2,7,8} 
C_i^{(0)\rm eff}\; C_j^{(0)\rm eff}\; K_{ij}^{(2)\rm rem}. \label{p22rem}
\ee
The ``$\simeq$'' sign is used above only because $i,j=3,4,5,6$ are skipped in
the sum.

\newsection{Interpolation in $m_c$ \label{sec:interpol}}

In the present section, we are going to estimate~ $P_2^{(2)\rm rem}$~ for~
$m_c < m_b/2$~ by performing a certain interpolation between its large-$m_c$
asymptotic form and an assumed value at $m_c=0$. In particular, we shall
assume that the large-$\beta_0$ approximation is accurate at $m_c=0$, which
may be argued for by recalling that no charm mass renormalization effects
arise at that point.  Since such an assumption is obviously a weak point of
the calculation, two alternative forms of it will be considered:
\mathindent0cm
\bea
{\rm (a)}&~~& P_2^{(2)\rm rem} ~\bbuildrel{\longrightarrow}_{z\to 0}^{}~ 0, \label{mc0.a}\\
{\rm (b)}&~~& P_1^{(2)} + P_2^{(2)} + P_3^{(2)} ~\bbuildrel{\longrightarrow}_{z\to 0}^{}~ 
P_2^{(2)\beta_0}. \label{mc0.b}
\eea
\mathindent1cm
The difference between these two cases will serve as a basis to estimate
  the interpolation uncertainty. As a cross-check, we shall also consider the 
  case 
\mathindent0cm
\bea
{\rm (c)}&~~& P_2^{(2)\rm rem} ~\bbuildrel{\longrightarrow}_{z\to 0}^{}~ 
\left( C_7^{(0)\rm eff} \right)^2 \widetilde{K}_{77}^{(2)\rm rem}(z=0), \label{mc0.c}
\eea
\mathindent1cm
which does not rely on the large-$\beta_0$ approximation but rather on
assuming that the ``77'' term dominates at $m_c=0$. In Appendix~B,
issues related to the $c\bar c$ production are discussed.

The functional form of $P_2^{(2)\rm rem}(z)$ that we are going to use for the
interpolation is a linear combination
\be \label{p2remxi}
P_2^{(2)\rm rem} = x_1\, |r_2^{(1)}(z)|^2 + x_2\, {\rm Re}\,r_2^{(2)}(z)
                    + x_3\, {\rm Re}\,r_2^{(1)}(z) + x_4\, z\f{d}{dz}{\rm Re}\,r_2^{(1)}(z) + x_5.
\ee
If $P_2^{(2)\rm rem}$ was dominated by renormalization effects, the last
three terms would be sufficient. The first two terms are included because
otherwise no $\ln^2 z$~ would be reproduced at large $z$.\linebreak The
function $|r_2^{(1)}|^2$ is non-analytic at~ $m_c = m_b/2$~ and takes into
account the pure four-quark operator contribution to the squared amplitude. The
function ${\rm Re}\,r_2^{(2)}(z)$ matches the remaining~ $\ln^2 z$~ in the
large-$m_c$ asymptotics. We use this function because it is the only genuine
NNLO four-quark operator matrix element that is known for small $z$.
\begin{figure}[t]
\begin{center}
\includegraphics[width=81mm,angle=0]{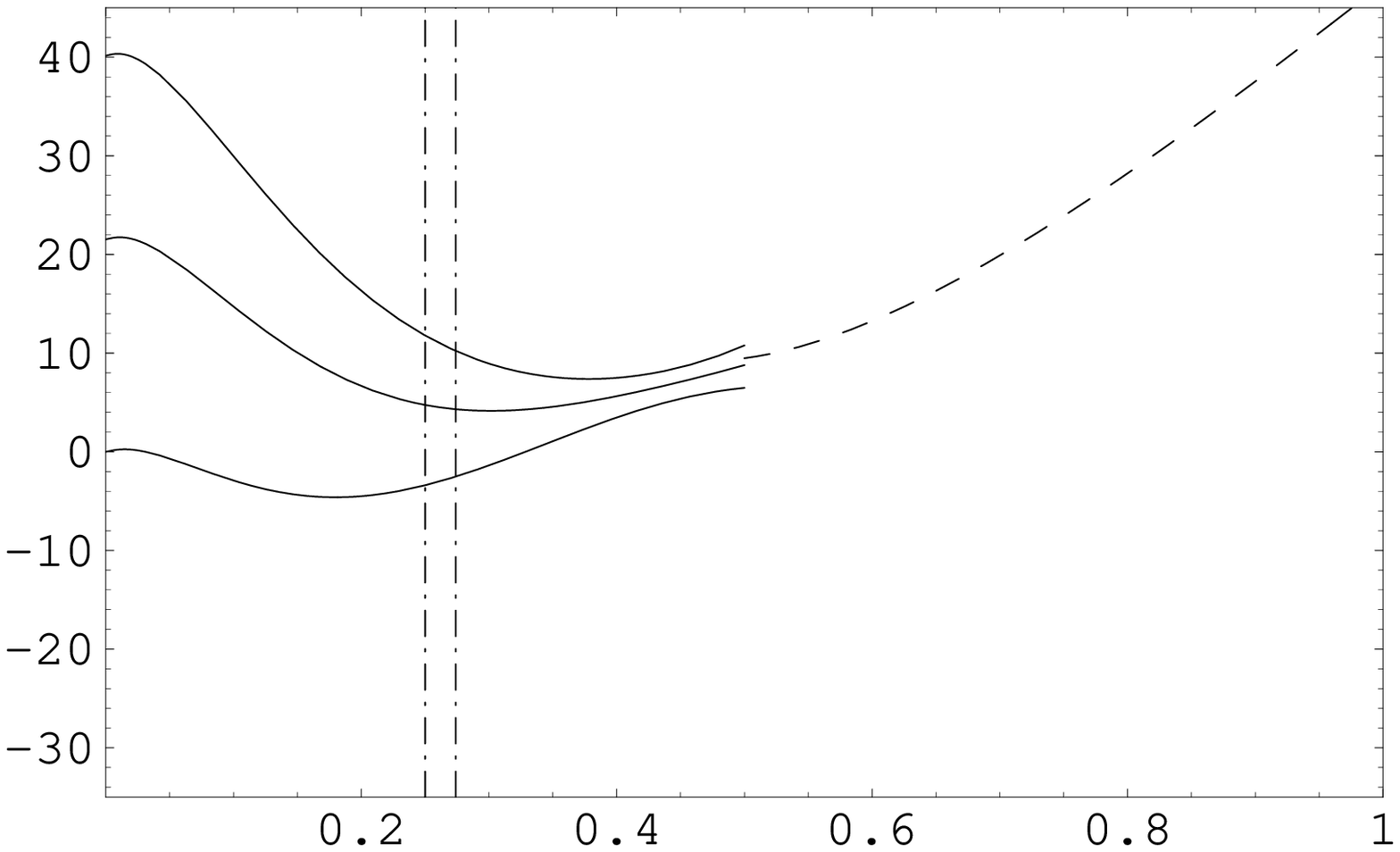}
\hspace{5mm}
\includegraphics[width=81mm,angle=0]{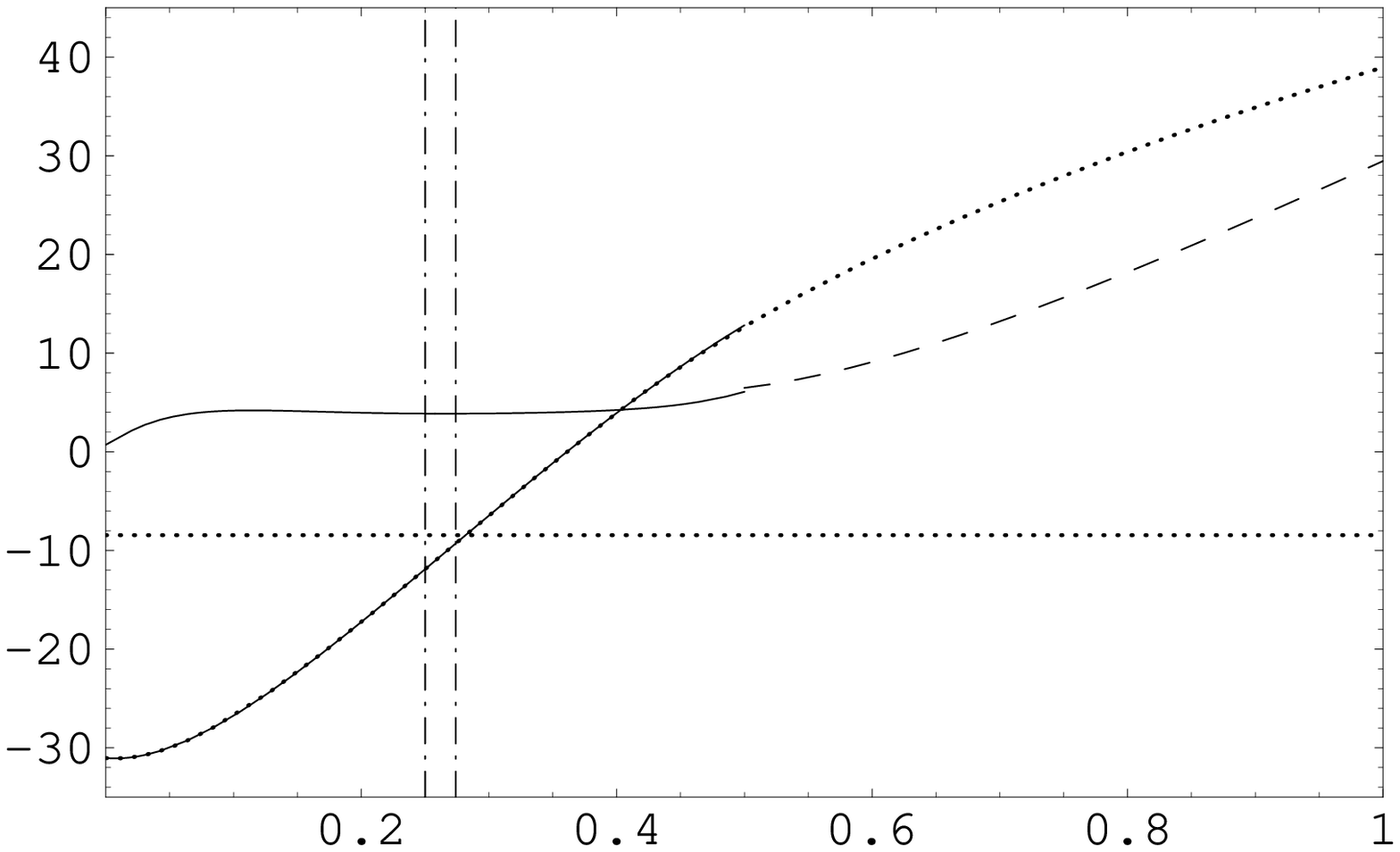} 
\end{center}
\vspace*{-51mm}
\hspace*{13mm} ${\scriptstyle\rm (b)}$
\hspace*{36mm} $P_2^{(2)\rm rem}$
\hspace*{74mm} $P_3^{(2)}$\\[6mm] 
\hspace*{9mm} ${\scriptstyle\rm (c)}$\\[-3mm]
\hspace*{10cm} $P_2^{(2)\beta_0}$\\[1.5mm]
\hspace*{7mm} ${\scriptstyle\rm (a)}$
\hspace*{137mm} $P_1^{(2)}$\\[1cm]
\hspace*{65mm} $m_c/m_b$ \hspace{73mm} $m_c/m_b$\\[-3mm]
\begin{center}
\caption{\sf $P_2^{(2)\rm rem}$, $P_2^{(2)\beta_0}$, $P_1^{(2)}$ and $P_3^{(2)}$ 
as functions of $m_c/m_b = \sqrt{z}$ for $\mu_0=2 M_W$,~ $\mu_b = m_b^{1S}/2$~ and~ $\mu_c = m_c(m_c)$.
The other input parameters are set to their central values from Appendix~A.
See the text below Eq.~(\ref{p32z}). \label{fig:pxx}}
\end{center}
\end{figure}
 
The determination of the coefficients $x_i$ is most easily explained in a
specific numerical example. Let us choose the renormalization scales as
  follows: $\mu_0=2 M_W$,~ $\mu_b = m_b^{1S}/2$~ and~ $\mu_c = m_c(m_c)$.
When the central values of the input parameters from Appendix~A are used but
only $m_c$ is retained arbitrary, Eq.~(\ref{p22rem}) yields
\be \label{p2rembig} 
P_2^{(2)\rm rem} \simeq (9.57 + 7.12) \ln^2 z + 50.52 \ln z +
47.44 + {\cal O}\left(\f{1}{z}\right), 
\ee
where the contribution from Eq.~(\ref{k22rem}) has been singled out in the
coefficient at~ $\ln^2 z$~(the first term). This first term is assumed to
determine $x_1$.  Then the second term in the coefficient at~ $\ln^2 z$~
determines $x_2$. Next, $x_3$ is determined from the coefficient at~ $\ln z$.~
Finally, by adjusting $x_4$ and $x_5$, one can simultaneously match the
constant term in Eq.~(\ref{p2rembig}) and satisfy one of the requirements
(\ref{mc0.a}), (\ref{mc0.b}) or (\ref{mc0.c}). In our example, the results are
\mathindent0cm
\bea
{\rm (a)}~~~ P_2^{(2)\rm rem}(z) &\simeq& 
   1.45 \left[ |r_2^{(1)}(z)|^2 - |r_2^{(1)}(0)|^2 \right]
+  8.01\,{\rm Re} \left[ r_2^{(2)}(z) - r_2^{(2)}(0) \right]
\nnb\\&+&
  15.63\,{\rm Re} \left[ r_2^{(1)}(z) - r_2^{(1)}(0) \right]
+ 16.52\, z\f{d}{dz}{\rm Re}\,r_2^{(1)}(z), \label{np2rema}\\
{\rm (b)}~~~ P_2^{(2)\rm rem}(z) &\simeq& 
   1.45 \left[ |r_2^{(1)}(z)|^2 - |r_2^{(1)}(0)|^2 \right]
+  8.01\,{\rm Re} \left[ r_2^{(2)}(z) - r_2^{(2)}(0) \right]
\nnb\\&+&
  15.63\,{\rm Re} \left[ r_2^{(1)}(z) - r_2^{(1)}(0) \right]
+ 0.89\, z\f{d}{dz}{\rm Re}\,r_2^{(1)}(z) + 40.15
, \label{np2remb}\\
{\rm (c)}~~~ P_2^{(2)\rm rem}(z) &\simeq& 
   1.45 \left[ |r_2^{(1)}(z)|^2 - |r_2^{(1)}(0)|^2 \right]
+  8.01\,{\rm Re} \left[ r_2^{(2)}(z) - r_2^{(2)}(0) \right]
\nnb\\&+&
  15.63\,{\rm Re} \left[ r_2^{(1)}(z) - r_2^{(1)}(0) \right]
+ 8.14\, z\f{d}{dz}{\rm Re}\,r_2^{(1)}(z) + 21.53, \label{np2remc}
\eea
\mathindent1cm
where $P_1^{(2)} + P_3^{(2)}(z=0) \simeq -40.15$
determines the very last term in the (b) case.

Substituting the central value of $m_c(m_c)$ from Appendix~A sets $z$ to $z_0
\simeq (0.262)^2 \simeq 0.0684$. Then,~ $P_2^{(2)\beta_0}(z_0) \simeq
3.86$~ and
\mathindent0cm
\bea
{\rm (a)}~~~ P_2^{(2)\rm rem}(z_0) \;\simeq\; -3.00
\hspace{1cm}  &\Rightarrow& \hspace{1cm} P_2^{(2)}(z_0) \;\simeq\; 0.87,\\[2mm]
{\rm (b)}~~~ P_2^{(2)\rm rem}(z_0) \;\simeq\; 10.98
\hspace{11mm} &\Rightarrow& \hspace{1cm} P_2^{(2)}(z_0) \;\simeq\; 14.85,\\[2mm]
{\rm (c)}~~~ P_2^{(2)\rm rem}(z_0) \;\simeq\; 4.50
\hspace{13mm} &\Rightarrow& \hspace{1cm} P_2^{(2)}(z_0) \;\simeq\; 8.36.
\eea
\mathindent1cm

The corrections $P_2^{(2)\rm rem}$ from Eqs.~(\ref{np2rema}), (\ref{np2remb})
and (\ref{np2remc}) are shown in the left plot of Fig.~\ref{fig:pxx}
as functions of $\sqrt{z}=m_c/m_b$.  The other NNLO corrections $P_1^{(2)}$,
$P_2^{(2)\beta_0}$ and $P_3^{(2)}$ are shown in the right plot. For
the same parameters, one finds\footnote{$\,$
In $P_3^{(2)}(z)$~(\ref{p32z}) and $P_2^{(1)}(z)$ (\ref{p21z}), we display only the 
$z$-dependence that is due to $a(z)$ and $b(z)$. Other $z$-dependent effects
that originate from $\phi_{ij}^{(1)}(\delta)$ are very small.}
\mathindent0cm
\bea
P_1^{(2)} &\simeq& -8.45,\\[2mm]
P_2^{(2)\beta_0}(z) &\simeq& 10.35 \,{\rm Re} \left[ r_2^{(2)}(z) - r_2^{(2)}(0) \right]\nnb\\[1mm]
                          &+& 9.57 \,{\rm Re} \left[ r_2^{(1)}(z) - r_2^{(1)}(0) \right]
+ 0.71 \hspace{15mm} \Rightarrow \hspace{2mm} P_2^{(2)\beta_0}(z_0) ~\simeq~ 3.86, \label{p22b0}\\[2mm]
P_3^{(2)}(z) &\simeq& 17.00\,{\rm Re}\, a(z) + 16.83\,{\rm Re}\, b(z) - 31.04
\hspace{4mm} \Rightarrow \hspace{5.5mm} P_3^{(2)}(z_0) ~\simeq~ -10.64. \label{p32z}
\eea
\mathindent1cm
The solid lines in Fig.~\ref{fig:pxx} show the small-$m_c$ expansions up to
${\cal O}(z^4)$. The dashed lines describe the leading terms in the
large-$m_c$ expansions. The dotted lines correspond to exact
  expressions. The (a), (b) and (c) cases of the interpolated
$P_2^{(2)\rm rem}$ are indicated in the plot. The two vertical 
  dash-dotted lines mark the $1\sigma$ range for $m_c(m_c)/m_b^{1S}$.
  
  As in Fig.~\ref{fig:abr2}, the large- and small-$m_c$ expansions nicely
  match at $m_c=m_b/2$. It is only a consequence of the properties of~
    $r_2^{(1)}$~ and~ ${\rm Re}\,r_2^{(2)}$~ that determine the $z$-dependence
    of all the considered functions.

It should be stressed that the shape of the curves in Fig.~\ref{fig:pxx} is
quite sensitive to the choice of renormalization scales. For instance, when the
charm mass renormalization scale $\mu_c$ is shifted from its default value
$\mu_c = m_c(m_c)$ to twice this value, the coefficients at~ 
$z\f{d}{dz}\,{\rm Re}\,r_2^{(1)}(z)$~ 
in Eqs.~(\ref{np2rema}), (\ref{np2remb}) and (\ref{np2remc}) change
quite dramatically (to $3.30$, $-12.34$ and $-5.09$, respectively).
However, the resulting effect on the decay rate is partly compensated
by a correlated change of $m_c(\mu_c)$ in the NLO correction
$P_2^{(1)}$. The numerical relevance of $P_2^{(2)\beta_0}$ is
very much $\mu_b$-dependent, which compensates the $\mu_b$-dependence
of $\alpha_s(\mu_b)$ in the NLO correction. The relatively small value
in Eq.~(\ref{p22b0}) indicates that $\mu_b = m_b^{1S}/2$ that is used
in this section is in the vicinity of the so-called BLM scale
\cite{Brodsky:1982gc}. The renormalization scale dependence of our
results will be discussed in more detail at the level of the branching
ratio in the next section.

From the above results for $P_k^{(2)}$ and
\mathindent0cm
\bea
P^{(0)}   &\simeq&   0.1396,\\
P_1^{(1)} &\simeq&  -1.515,\\
P_2^{(1)}(z) &\simeq&  3.347 - 1.826\,{\rm Re}\, a(z) - 1.652\,{\rm Re}\, b(z)
\hspace{7mm} \Rightarrow \hspace{7mm} P_2^{(1)}(z_0) \simeq 1.151, \label{p21z}
\eea
\mathindent1cm
that are calculated using the same parameters, we obtain
\mathindent0cm
\bea
{\rm (a)}~~~ P(E_0) &\simeq& 0.1226 + {\cal O}\left(\alpha_{\rm em}, \alpha_s V_{ub}\right)
                    ~\simeq~ 0.1192,\\[2mm]
{\rm (b)}~~~ P(E_0) &\simeq& 0.1295 + {\cal O}\left(\alpha_{\rm em}, \alpha_s V_{ub}\right)
                    ~\simeq~ 0.1261,\\[2mm]
{\rm (c)}~~~ P(E_0) &\simeq& 0.1263 + {\cal O}\left(\alpha_{\rm em}, \alpha_s V_{ub}\right)
                    ~\simeq~ 0.1229.
\eea
\mathindent1cm
In the second step above, the electroweak and ${\cal O}\left(V_{ub}\right)$
corrections are added according to Eq.~(3.10) of 
Ref.~\cite{Gambino:2001au}\footnote{$\,$
We are grateful to the authors of Ref.~\cite{Gambino:2001au} for
    providing us with their results in an extended version that allows
    for arbitrarily varying $\mu_0$ and $\mu_b$.}
and Eq.~(3.7) of Ref.~\cite{Gambino:2001ew}, respectively. 

Next, using Eq.~(\ref{main}) with $N(E_0) \simeq 0.0031$, one finds
\mathindent0cm
\bea
{\rm (a)}~~~  {\cal B}[\bar{B} \to X_s \gamma]_{\scs E_{\gamma} > 1.6\,{\rm GeV}}
&\simeq& 3.02 \times 10^{-4},\label{bra}\\[2mm]
{\rm (b)}~~~  {\cal B}[\bar{B} \to X_s \gamma]_{\scs E_{\gamma} > 1.6\,{\rm GeV}}
&\simeq& 3.19 \times 10^{-4},\label{brb}\\[2mm]
{\rm (c)}~~~  {\cal B}[\bar{B} \to X_s \gamma]_{\scs E_{\gamma} > 1.6\,{\rm GeV}}
&\simeq& 3.11 \times 10^{-4}.\label{brc}
\eea
\mathindent1cm

\newsection{Estimating the uncertainties\label{sec:uncert}}

\begin{figure}[t]
\begin{center}
\includegraphics[width=8cm,angle=0]{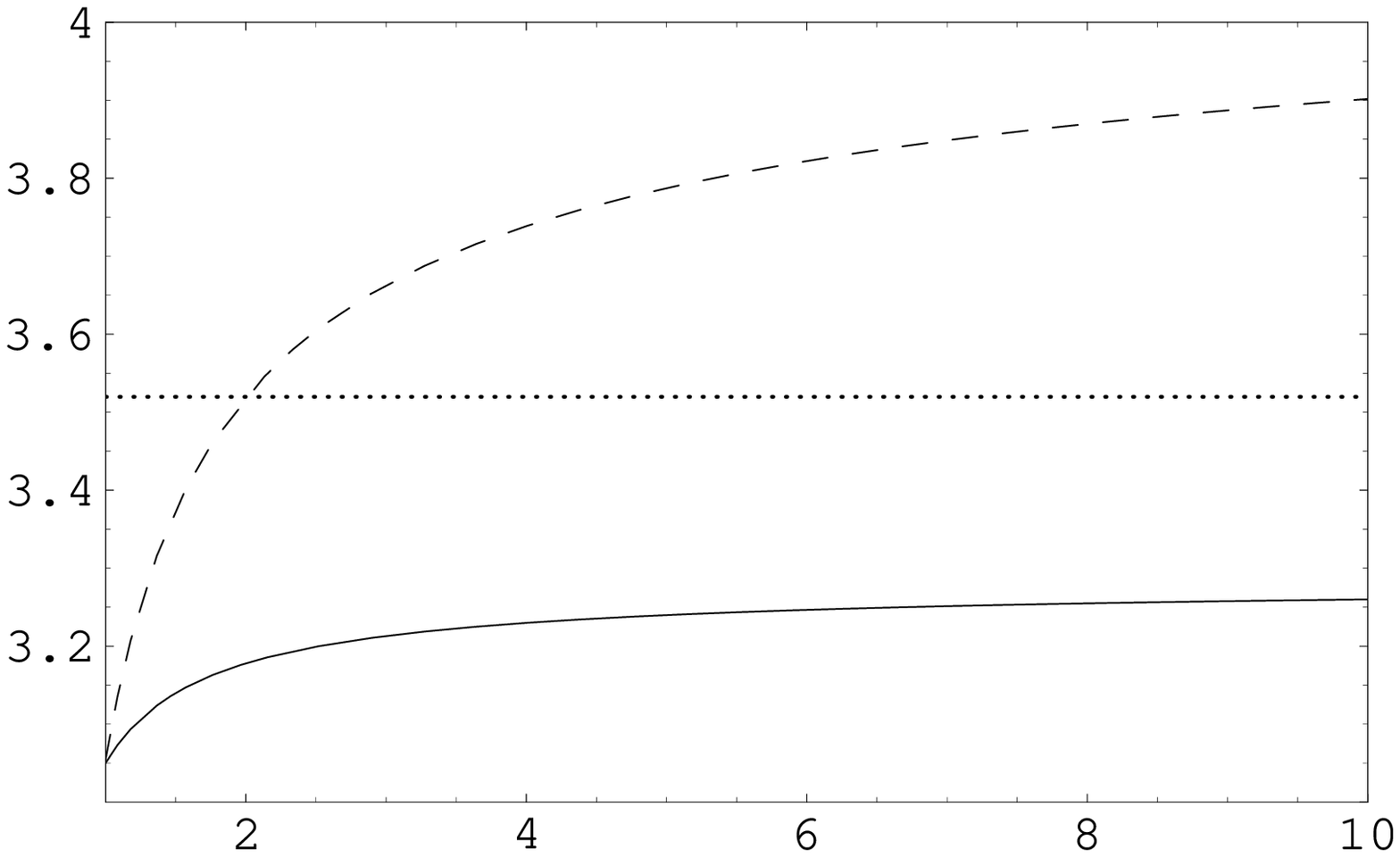}
\hspace{5mm}
\includegraphics[width=8cm,angle=0]{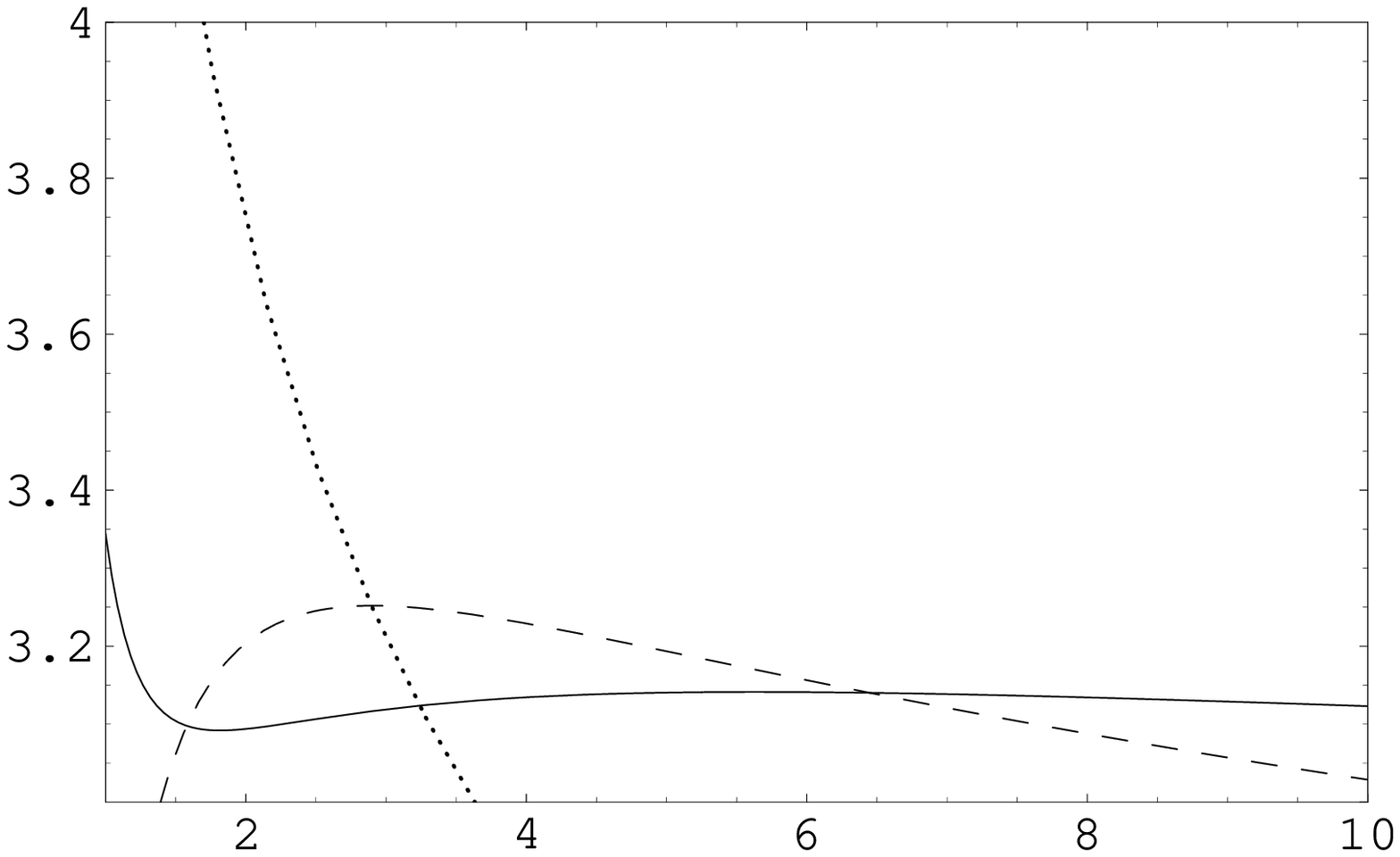}\\
\includegraphics[width=8cm,angle=0]{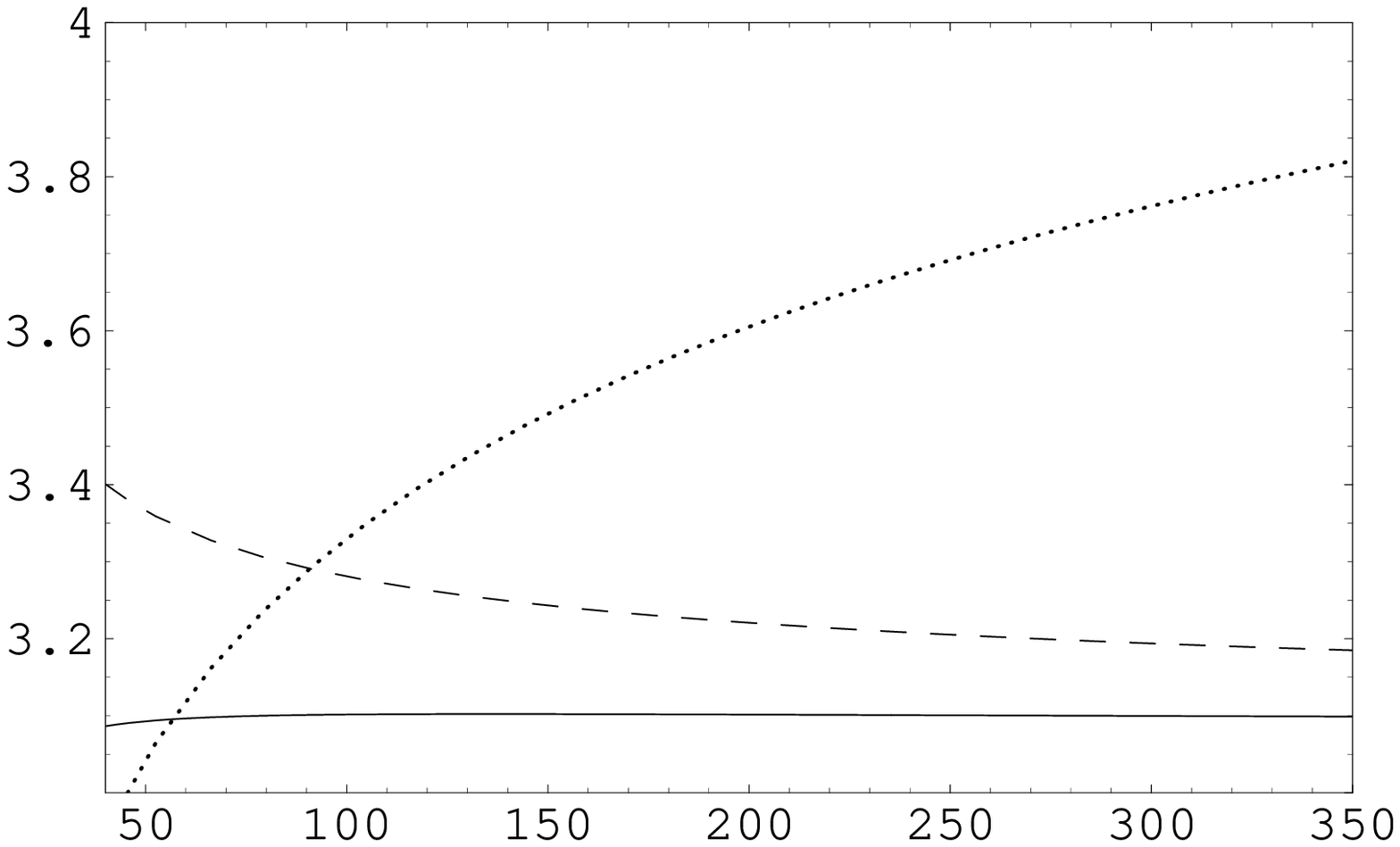}
\end{center}
\vspace*{-67mm} 
\hspace*{39mm} $\mu_c$ \hspace{81mm} $\mu_b$\\[46mm]
\hspace*{86mm} $\mu_0$\\ 
\begin{center}
\caption{\sf Renormalization scale dependence of ${\cal B}(\bar{B} \to X_s \gamma)$
  in units $10^{-4}$ at the LO (dotted lines), NLO (dashed lines) and NNLO (solid
  lines). The upper-left, upper-right and lower plots describe the dependence
  on $\mu_c$, $\mu_b$ and $\mu_0$ [GeV], respectively. \label{fig:mudep}}
\end{center}
\end{figure}

The results (a) and (b) given at the end of the previous section are
  within the range of around $\pm 3\%$ from their average, and the result (c)
  is in between them. The spread of the three solutions decreases for
  higher scales $\mu_b$. Moreover, the left plot in Fig.~\ref{fig:pxx} seems
  reasonable even as a description of an extrapolation rather than an
  ``interpolation with an assumption''. Therefore, we shall assign a fixed
  $\pm 3\%$ to the interpolation ambiguity.\footnote{$\,$
  Various uncertainties will be added in quadrature later. Thus, each of
    them should be understood as a ``theoretical 1$\sigma$'' rather than 
    a strict range. Of course, such statements concerning theoretical 
    uncertainties are never well-defined, but one can hardly improve on this point.}
  Unfortunately, such a rough way of determining this error may need to remain
  in the $\bar{B} \to X_s \gamma$ analysis until a complete evaluation of the
  three-loop matrix elements is performed. Calculating them just in the
  $m_c=0$ case would help a lot.
  
As far as the higher-order (${\cal O}(\alpha_s^3)$) perturbative uncertainties
are concerned, renormalization-scale dependence is usually used to place lower
bounds on their size. The three plots in Fig.~\ref{fig:mudep} show the
branching ratio dependence on each of the three scales, once the remaining two are
fixed at the values that were used in the previous section ($\mu_0=2 M_W$,~
$\mu_b = m_b^{1S}/2$~ and~ $\mu_c = m_c(m_c)$).  Dotted, dashed and solid
lines describe the LO, NLO and NNLO results.  The NNLO branching ratio is
defined as the average of the (a) and (b) cases.  Stabilization of the
scale dependence with growing order in $\alpha_s$ is clearly seen. It is
especially encouraging in the case of $\mu_b$ because the interpolation
assumptions (\ref{mc0.a}) and (\ref{mc0.b}) violate the analytic
$\mu_b$-dependence cancellation even at ${\cal O}(\alpha_s^2)$.  The
$\mu_c$-stabilization is more obvious because it is built into our
interpolation prescription by construction. The perfect $\mu_0$-stabilization
arises both due to the smaller value of $\alpha_s$ at the matching scale and
to the fact that the NNLO matching conditions are complete.

  By studying plots like those in Fig.~\ref{fig:mudep} for various choices of the fixed
  scales, we have convinced ourselves that $\pm 3\%$ is a reasonable estimate
  of the higher order perturbative uncertainty. At the same time, we find
  that $3.15 \times 10^{-4}$ is a good central value. It is reproduced
  (as an average of the (a) and (b) cases), e.g., for 
  $\mu_0= 160\;{\rm GeV}$,~ $\mu_b = 2.5\;{\rm GeV}$~ and~ $\mu_c = 1.5\;{\rm GeV}$. 
  Table~\ref{tab:abcmubc} contains our results for the 
  branching ratio in various cases and for several particular choices of the
  renormalization scales.

\begin{table}[t]
\begin{center}
\begin{tabular}{|c||c|c|c||c|c|c|}\hline
$\mu_b\;$[GeV] & 2 & 2.5 & 5 & \multicolumn{2}{|c|}{2.5} \\\hline 
$\mu_c\;$[GeV] & \multicolumn{3}{|c||}{1.5} & $m_c(m_c)$ & 2.5 \\\hline
(a)            & 3.01 & 3.06 & 3.14 & 3.03 & 3.10 \\
(b)            & 3.23 & 3.24 & 3.26 & 3.19 & 3.33 \\
(c)            & 3.16 & 3.15 & 3.18 & 3.10 & 3.21 \\
\ [(a)+(b)]/2  & 3.12 & 3.15 & 3.20 & 3.11 & 3.22 \\\hline
\end{tabular}
\caption{\sf ${\cal B}(\bar{B} \to X_s \gamma) \times 10^4$ 
              in various cases for several choices of $\mu_b$ and $\mu_c$.
              The matching scale $\mu_0$ is set to $160\;{\rm GeV}$,
              and the remaining parameters are as in Appendix A. \label{tab:abcmubc}} 
\end{center}
\end{table}
  The parametric uncertainty is the most straightforward one. Our input
    parameters and their uncertainties are listed in Appendix~A.
    The correlation between $C$ and $m_c(m_c)$ is taken into
    account. The remaining errors that are quoted in the right columns
    of Tables~\ref{tab:inp.overall}, \ref{tab:inp.pp} and~\ref{tab:inp.remaining}
    in Appendix~A are treated as uncorrelated. This way we find the overall
    parametric uncertainty of $\pm 3.0\%$.
  
  Finally, the non-perturbative uncertainties need to be considered. It has
  been customary for a long time to quote the known 
  ${\cal O}(\Lambda^2/m_b^2)$ and ${\cal O}(\Lambda^2/m_c^2)$ non-perturbative
  corrections as the dominant ones (see Section~VI of Ref.~\cite{Hurth:2003vb}
  for discussion and references). The subdominant 
  ${\cal O}(\Lambda^2/m_b^3)$ and ${\cal O}(\Lambda^2/(m_c^2 m_b))$ ones
  are known, too~\cite{Bauer:1997fe}. More recently, the ${\cal O}(\alpha_s
  \Lambda^2/(m_b-2 E_0)^2)$ correction was evaluated~\cite{Neubert:2004dd}.
  All these corrections are included in $N(E_0)$ in Eq.~(\ref{main}) and
    cause around 2.4\% enhancement of the branching ratio. However, the
  non-perturbative corrections that arise in the matrix elements of $Q_{1,2}$
  in the presence of one gluon that is not soft remain unknown. They scale like~
  $\alpha_s \Lambda/m_b$~ in the limit~ $m_c \ll m_b/2$~and like~
  $\alpha_s \Lambda^2/m_c^2$~ in the limit~ $m_c \gg m_b/2$.\footnote{$\,$
We thank M.~Beneke for a clarifying discussion on this point.} 
Since $m_c < m_b/2$ in reality, we consider $\alpha_s \Lambda/m_b$~ as
the quantity that sets the size of such effects. In consequence, we assign a
$\pm 5\%$ non-perturbative uncertainty to our result. This is the dominant
uncertainty at the moment. The very recent estimates~\cite{Lee:2006wn} of
  similar corrections to the $Q_7$-$Q_8$ interference term are neglected here.
  They are smaller than the overall uncertainty of $\pm 5\%$ that both the
  authors of Ref.~\cite{Lee:2006wn} and us assign to {\em all} the unknown
  ${\cal O}(\alpha_s \Lambda/m_b)$ effects.

\newsection{ Conclusions \label{sec:conclusions}}

Our final NNLO result
\be
{\cal B}[\bar{B} \to X_s \gamma]_{\scs E_{\gamma} > 1.6\,{\rm GeV}}
~\simeq~ \left( 3.15 \pm 0.23  \right) \times 10^{-4}\label{br.final}
\ee
is obtained for the input parameters listed in Appendix~A and for the
  renormalization scales $\mu_0= 160\;{\rm GeV}$,~ $\mu_b = 2.5\;{\rm GeV}$~
  and~ $\mu_c = 1.5\;{\rm GeV}$. The total uncertainty is found by combining
  in quadrature the ones discussed in Section~\ref{sec:uncert}. As it is
  clearly seen in Fig.~\ref{fig:mudep}, the NNLO corrections significantly
  improve the stability of the prediction with respect to the renormalization
  scale variation and, in consequence, reduce the total error. Comparing the
  NLO and NNLO $\mu_c$-dependence in Fig.~\ref{fig:mudep}, one realizes why
  all the previously published NLO predictions had significantly higher
  central values than the current NNLO one.
  
  In order to relate our result with $E_0=1.6\,$GeV to the measurements with
  cuts at $1.8\,$GeV (Belle~\cite{Koppenburg:2004fz}) and $1.9\,$GeV
  (BaBar~\cite{Aubert:2006gg}), one needs to compute ratios of the
  decay rates with different cuts (see, e.g., Ref.~\cite{Buchmuller:2005zv}).  
  This is a non-trivial issue because new perturbative and non-perturbative effects 
  become important in the endpoint region. A new calculation of such effects 
  has recently been completed~\cite{Becher:2005pd} but its numerical 
  results were not yet available when the average in Eq.~(\ref{eq:HFAG}) 
  was being evaluated.

We have chosen $E_0 = 1.6\;{\rm GeV}$ as default here assuming that it is low enough
for the cutoff-enhanced perturbative corrections~\cite{Neubert:2004dd,Becher:2005pd} 
to become negligible. If this turns out not to be the case, one should use our
results with lower $E_0$. For this purpose, we give\footnote{
These results do not include contributions from tree-level diagrams with
four quark operator insertions that are suppressed either by $|V_{ub}|^2$ or by
squares of the small Wilson coefficients $C_3,\ldots,C_6$\cite{Ligeti:1999ea}.}
\bea
{\cal B}[\bar{B} \to X_s \gamma]_{\scs E_{\gamma} > 1.5\,{\rm GeV}}
&\simeq& \left( 3.18 \pm 0.23  \right) \times 10^{-4},\label{br.1.5}\\[1mm]
{\cal B}[\bar{B} \to X_s \gamma]_{\scs E_{\gamma} > 1.4\,{\rm GeV}}
&\simeq& \left( 3.20 \pm 0.23  \right) \times 10^{-4},\label{br.1.4}\\[1mm]
{\cal B}[\bar{B} \to X_s \gamma]_{\scs E_{\gamma} > 1.3\,{\rm GeV}}
&\simeq& \left( 3.22 \pm 0.23  \right) \times 10^{-4},\label{br.1.3}\\[1mm]
{\cal B}[\bar{B} \to X_s \gamma]_{\scs E_{\gamma} > 1.2\,{\rm GeV}}
&\simeq& \left( 3.24 \pm 0.23  \right) \times 10^{-4},\label{br.1.2}\\[1mm]
{\cal B}[\bar{B} \to X_s \gamma]_{\scs E_{\gamma} > 1.1\,{\rm GeV}}
&\simeq& \left( 3.25 \pm 0.23  \right) \times 10^{-4},\label{br.1.1}\\[1mm]
{\cal B}[\bar{B} \to X_s \gamma]_{\scs E_{\gamma} > 1.0\,{\rm GeV}}
&\simeq& \left( 3.27 \pm 0.23  \right) \times 10^{-4}.\label{br.1.0}
\eea

\section*{Note added}

A numerical analysis of cutoff-related perturbative corrections (see
  Section~\ref{sec:conclusions}) became available \cite{Becher:2006pu} when
  the current paper was being refereed. The authors of
  Ref.~\cite{Becher:2006pu} use our result given in Eq.~(\ref{br.1.0}) and
  combine it with their cutoff-related corrections that turn out to suppress
  the branching ratio at $E_0=1.6\;{\rm GeV}$ by around 3\% with respect to
  Eq.~(\ref{br.final}). Such an effect is of the same size as our higher-order
  uncertainty, which means that the unknown $O(\alpha_s^3)$ non-logarithmic
  effects can be as important at $E_0=1.6\;{\rm GeV}$ as the logarithmic ones
  calculated in Ref.~\cite{Becher:2006pu}. Therefore, we leave the results of
  the current paper unaltered, yet not excluding the possibility of taking the
  correction from Ref.~\cite{Becher:2006pu} into account in future upgrades of
  the phenomenological analysis, once other effects of potentially the
  same size are collected (see Section 1 of Ref.~\cite{Asatrian:2006rq}). 

\section*{Acknowledgments}

We would like to thank Aneesh Manohar for information on correlations among
parameters that are determined in the global semileptonic fit in
Refs.~\cite{Bauer:2002sh,Bauer:2004ve,Hoang:2005zw}.  We are grateful to
Paolo Gambino and Ulrich Haisch for cross-checking the numerical results for
$P_3^{(2)}$ and $P_1^{(2)}$, respectively, as well as for their help
concerning the ${\cal O}(\Lambda^3)$ and electroweak corrections.
This work was supported in part by the EU Contract MRTN-CT-2006-035482, FLAVIAnet.
M.M. acknowledges support by the Polish Committee for Scientific Research
under the grant 2~P03B~078~26, and from the EU Contract 
HPRN-CT-2002-00311, EURIDICE.  M.S. acknowledges support by the DFG through SFB/TR~9.

\newappendix{Appendix~A:~~ Numerical inputs}
\def\theequation{A.\arabic{equation}}

In this appendix, we collect numerical parameters that are used in
Sections~\ref{sec:interpol} and \ref{sec:uncert}. The default value of the
photon energy cut is $E_0 = 1.6\,{\rm GeV}$. In
  Section~\ref{sec:interpol}, the renormalization scales were~ $\mu_0=2 M_W$,~
  $\mu_b = m_b^{1S}/2$~ and~ $\mu_c = m_c(m_c)$. The final central value of
  the branching ratio in Eq.~(\ref{br.final}) is reproduced e.g. for~ $\mu_0=
  160\;{\rm GeV}$,~ $\mu_b = 2.5\;{\rm GeV}$~ and~ $\mu_c = 1.5\;{\rm
    GeV}$.

The other parameters are displayed in the tables below.  Errors are indicated
only if varying a given parameter within its $1\sigma$ range causes a larger
than $\pm 0.1\%$ effect on the branching ratio (\ref{br.final}). In such
cases, the effects in percent are given in the right column of the
corresponding table.  

Table~\ref{tab:inp.overall} contains the four quantities that determine the
overall normalization factor multiplying $P(E_0)$ in the expression
(\ref{main}) for the branching ratio.

Table~\ref{tab:inp.pp} contains the experimental inputs that are necessary in
the calculation of $P(E_0)$, before including the electroweak and ${\cal
  O}(V_{ub})$ corrections. For $\alpha_s(M_Z)$, we adopt the central value
from Ref.~\cite{Bethke:2006ac} but conservatively use a twice larger error
which then overlaps with the one given by the PDG \cite{Yao:2006px}.  One should
take into account that the phase space factor $C$ in
Table~\ref{tab:inp.overall} and $m_c(m_c)$ in Table~\ref{tab:inp.pp} are
strongly correlated. For this reason, we take both parameters as determined in
the same global fit to the semileptonic $B$-decay spectra
\cite{Bauer:2002sh,Bauer:2004ve,Hoang:2005zw} instead of adopting $m_c(m_c)$ from
Ref.~\cite{Kuhn:2001dm}. The normalized correlation coefficient amounts
to~\cite{Manohar:2006priv}
\be
F \equiv \f{ \langle C\, m_c \rangle - \langle C \rangle \langle m_c \rangle }{
        \sigma_{\scs C} \sigma_{m_c} }
\simeq \left\{ \begin{array}{cc} -0.97, & \mbox{~~method A},\\[1mm] 
                                 -0.92, & \mbox{~~method B},\end{array}\right.
\ee
where the meaning of the two methods is explained in Section III of
Ref.~\cite{Hoang:2005zw}. Denoting the uncertainties in the right columns of
Tables~\ref{tab:inp.overall} and~\ref{tab:inp.pp} by $\Delta_x$, where $x$ is
the relevant variable, one finds that the combined uncertainty due to $C$ and
$m_c$ is given by
\be
\Delta_{C,m_c} = \sqrt{ \Delta^2_C + \Delta^2_{m_c} + 2 F |\Delta_C \Delta_{m_c}| }.
\ee
Since $\Delta_C$ and $\Delta_{m_c}$ are very close in size (by coincidence)
and $F$ is close to $-1$, the combined uncertainty is much smaller than it
would be in the absence of the correlation. For our final result, we adopt
$F=-0.92$ that gives larger $\Delta_{C,m_c}$ ($\pm 1.1\%$ rather than $\pm
0.7\%$).

\begin{table}[t]
\begin{center}
\begin{tabular}{|l|l|}
\hline
parameter & effect on Eq.~(\ref{br.final})\\[1mm]\hline
  ${\cal B}(B\to X_c e \bar\nu)_{\rm exp} = 0.1061 \pm 0.0017$~\cite{Aubert:2004aw}
& $\pm 1.6\%$  \\[1mm]
  $C = 0.580 \pm 0.016$~\cite{Bauer:2004ve,Manohar:2006priv} 
& $\pm 2.8\%$  \\[1mm]
  $|V_{ts}^* V_{tb}/V_{cb}|^2 = 0.9676 \pm 0.0033$~\cite{Bona:2006ah,Charles:2004jd}
& $\pm 0.4\%$  \\[1mm]
  $\alpha_{\rm em}(0) =  1/137.036$~\cite{Yao:2006px}
& \\\hline
\end{tabular}
\caption{\sf Parameters that determine the overall normalization factor in Eq.~(\ref{main}).
             \label{tab:inp.overall}} 
\end{center}
\end{table}
\begin{table}[t]
\begin{center}
\begin{tabular}{|l|l|}
\hline
  parameter & effect on Eq.~(\ref{br.final})\\[1mm]\hline
  $M_Z = 91.1876\;{\rm GeV}$~\cite{Yao:2006px}
& \\[1mm]
  $\alpha_s (M_Z) = 0.1189 \pm 0.0020$~\cite{Bethke:2006ac,Yao:2006px}
& $\pm 2.0\%$ \\[1mm]
  $m_{t,{\rm pole}}= (171.4 \pm 2.1) \;{\rm GeV}$~\cite{Group:2006xn}
& $\pm 0.5$\% \\[1mm]
  $M_W = 80.403\;{\rm GeV}$~\cite{Yao:2006px}
& \\[1mm]
  $m_b^{1S} = (4.68 \pm 0.03)\;{\rm GeV}$~\cite{Bauer:2004ve} 
& $\pm 0.2$\% \\[1mm]
  $m_c(m_c) = (1.224 \pm 0.017 \pm 0.054)\;{\rm GeV}$~\cite{Hoang:2005zw}  
& $\pm 2.8$\% \\\hline
\end{tabular}
\caption{\sf Experimental inputs that are
  necessary in the calculation of $P(E_0)$ before including
  the electroweak and ${\cal O}(V_{ub})$ corrections. \label{tab:inp.pp}}
\end{center}
\end{table}

The dependence of $m_c(m_c)$ on $\alpha_s(M_Z)$ in the semileptonic fit can
affect $m_c(m_c)$ by up to~ $\sim 10\;$MeV \cite{Manohar:2006priv}, which
would translate to~ $\sim 0.5\%$ in $\Delta_{\alpha_s}$. We neglect this
effect here. An analogous correlation for $C$ is also neglected, as it turns
out to be even smaller \cite{Manohar:2006priv}. 
\begin{table}[t]
\begin{center}
\begin{tabular}{|l|l|}
\hline
  parameter & effect on Eq.~(\ref{br.final})\\[1mm]\hline
  $\alpha_{\rm em}(M_Z) =  1/128.940$ \cite{Kuhn:1998ze,Burkhardt:2005se}
&  \\[1mm] 
  $\sin^2\theta_W = 0.2324$ \cite{LEPEWWG.2006}
&  \\[1mm]                              
  $M_{\rm Higgs} \in [114.4, 194]\;{\rm GeV}$ (95\% C.L)~\cite{Yao:2006px}
& $0.3\%$ \\[1mm] 
  \hspace{15mm} ($115\;{\rm GeV}$ is used as central) &\\[1mm]
  $ \left(V_{us}^* V_{ub}\right)/\left(V_{ts}^* V_{tb}\right) 
    = -0.011 + 0.018 i$~\cite{Bona:2006ah,Charles:2004jd}
& \\[1mm] 
  $\lambda_1 = (-0.27 \pm 0.04) \;{\rm GeV}^2$~\cite{Bauer:2004ve}
& \\[1mm]
  $\lambda_2 \simeq \f{1}{4} \left(m_{B^*}^2-m_B^2\right) \simeq 0.12 \;{\rm GeV}^2$~\cite{Yao:2006px}
& \\[1mm] 
  $\rho_1 = (0.038 \pm 0.028) \;{\rm GeV}^3$~\cite{Bauer:2004ve,Manohar:2006priv} 
& $\pm 0.4\%$\\[1mm]
  $\rho_2 = (0.0045 \pm 0.035) \;{\rm GeV}^3$~\cite{Bauer:2004ve,Manohar:2006priv} 
& $\pm 0.6\%$ \\\hline
\end{tabular}
\caption{\sf Remaining parameters that are necessary for the electroweak, ${\cal O}(V_{ub})$ 
  and non-perturbative corrections.
             \label{tab:inp.remaining}} 
\end{center}
\end{table}
\begin{table}[ht]
\begin{center}
\begin{tabular}{|r|r|r|}
\hline
$i$ & $C_i^{(0)\rm eff}(\mu_b)$ & $C_i^{(1)\rm eff}(\mu_b)$\\\hline
1 & $-$0.8411 &   15.278\\
2 &    1.0647 & $-$2.124\\
3 & $-$0.0133 &    0.096\\
4 & $-$0.1276 & $-$0.463\\
5 &    0.0012 & $-$0.021\\
6 &    0.0028 & $-$0.013\\
7 & $-$0.3736 &    2.027\\
8 & $-$0.1729 & $-$0.614\\
\hline
\end{tabular}
\caption{\sf The LO and NLO Wilson coefficients~$C_i^{(k)\rm eff}(\mu_b)$~ 
at $\mu_b=m_b^{1S}/2 = 2.34\;{\rm GeV}$. The matching scale $\mu_0$
was set to $2 M_W$ in their evaluation.\label{tab:Wilson}} 
\end{center}
\end{table}

Table~\ref{tab:inp.remaining} contains the remaining parameters that are
necessary for the electroweak, ${\cal O}(V_{ub})$ and non-perturbative
corrections. Since we treat the phase space factor $C$ as an independent
  input, $\lambda_1$ is needed only for the small cutoff-related
  non-perturbative correction~\cite{Neubert:2004dd}.
  
  The parameters $\rho_1$ and $\rho_2$ are needed for the ${\cal
      O}\left(\Lambda^3/m_b^3,\;\Lambda^3/(m_b m_c^2)\right)$ contributions to
    $N(E_0)$ that are derived from the formulae of
    Refs.~\cite{Bauer:1997fe,Gremm:1996df}.  An important subtlety in this
    calculation is that the ${\cal O}\left(\Lambda^3/m_b^3\right)$ corrections
    to $\Gamma\left(\bar{B} \to X_u e \nu\right)$ logarithmically diverge when
    $m_u \to 0$, so long as hard gluon interactions with the spectator quark are
    neglected. When they are included, one encounters new non-perturbative
    matrix elements whose values are rather uncertain~\cite{Gambino:2005tp}.
    Since these effects cancel in the ratio $(P(E_0)+N(E_0))/C$, we follow the
    approach that has been used in the calculation of $C$ in
    Ref.~\cite{Bauer:2004ve}, namely we neglect the spectator effects and set
    $\ln m_b/m_u$ to zero.  At the same time, we neglect the additional
    uncertainty in $C$ that is caused by this arbitrary procedure (see the
    comments below Eq.~(25) in Ref.~\cite{Bauer:2004ve}). Our final result
    (\ref{br.final}) is the same as if the considered subtlety did not
      occur at all. In the future, simultaneous calculations of $C$ and $N(E_0)$ 
     should be performed on the basis of the semileptonic fit results of 
     Ref.~\cite{Buchmuller:2005zv}, which would constitute an independent
     test of the the current calculation.
  
  Although the Wilson coefficients $C_i^{\rm eff}(\mu_b)$ are derived
  quantities, we quote for convenience their values at $\mu_b =
    m_b^{1S}/2$. The LO and NLO ones are collected in
  Table~\ref{tab:Wilson}.  From among the NNLO ones, only $C_7^{(2)\rm
      eff}(\mu_b)$ is relevant here. We find $C_7^{(2)\rm eff}(m_b^{1S}/2)
    \simeq 16.81$ including the four-loop mixing $(Q_1,\ldots,Q_6) \to Q_7$
    but neglecting the four-loop mixing $(Q_1,\ldots,Q_6) \to Q_8$.
  
The analytical solutions to the NNLO RGE for the Wilson coefficients can 
     be found, e.g., in Section~3.3 of Ref.~\cite{Huber:2005ig}. The resulting
     explicit expressions for $C_i^{(k)\rm eff}(\mu_b)$ in terms of  
     $C_i^{(k)\rm eff}(\mu_0)$ and $\eta = \alpha_s(\mu_0)/\alpha_s(\mu_b)$
     are given in Ref.~\cite{Czakon:2006ss}. As far as
     $\alpha_s(\mu)$ is concerned, we have used the four-loop RGE
     and applied two independent methods for solving it: a numerical one
     implemented in {\tt RunDec}~\cite{Chetyrkin:2000yt} and an iterative
     analytical one that includes (tiny) QED effects, too~\cite{Huber:2005ig}. 
     Our final results are independent of which method is used. However, the 
     intermediate quantities, for which more digits are presented, slightly 
     depend on the method.

\newappendix{Appendix~B:~~ The $m_c=0$ case and $c\bar c$ production}
\def\theequation{B.\arabic{equation}}

The ${\bar B}\to X_s\gamma$ branching ratio contains no contribution from
  $c\bar c$ production because events involving charmed hadrons in the
final state are not included on the experimental side.  Thus, the
evaluation of~ $b \to X_s^{\rm parton} \gamma$~ should be performed
accordingly.  However, such a definition of~ $b \to X_s^{\rm parton} \gamma$
may break down for $m_c\to 0$, since logarithmic divergences containing $\ln
m_c$ may arise in $P_2^{(2)\rm rem}$.
  
No such divergences are present in the NLO contributions to $P(E_0)$, and
  in the other than $P_2^{(2)\rm rem}$ NNLO ones. There, all the logarithms
of $m_c$ get multiplied by positive powers of this mass. If $P_2^{(2)\rm rem}$
turns out to be logarithmically divergent at $m_c\to 0$, it should be
redefined to include the $c\bar c$ production contributions. The assumptions
(\ref{mc0.a})--(\ref{mc0.c}) would then refer to the redefined quantity that
must be convergent at $m_c\to 0$. At the same time, the large-$m_c$
asymptotics would remain unaltered. After performing the interpolation of the
redefined quantity as in Section~\ref{sec:interpol}, one should subtract the
$c\bar c$ production effects at the measured value of $m_c$. At present,
possible effects of such a subtraction are understood to be contained in the
interpolation ambiguity.

Actually, Eq.~(\ref{k772remmc0}) does contain the $c\bar c$ production
contributions in the $m_c\to 0$ case. However, it is used only in the
assumption (\ref{mc0.c}) that serves just as a cross check, and has no
influence on our final numerical result (\ref{br.final}).

\setlength {\baselineskip}{0.2in}
 
\end{document}